\documentclass[
preprint, showkeys, 
amsmath,amssymb,
 aps, prb, 
 superscriptaddress, 
]{revtex4-2}

\usepackage{graphicx}
\usepackage{bm}
\usepackage{gensymb}

\usepackage{hyperref} 
\hypersetup{
  colorlinks = true,
  urlcolor   = blue,
  linkcolor  = blue,
  citecolor  = blue
}

\begin{document}

\title{ Investigation of the intrinsic hidden spin texture and spin-state segregation in centrosymmetric monolayer dichalcogenide: effectiveness of the electric-field approach}

\author{Ameneh Deljouifar}
\affiliation{Computational Nanophysics Laboratory (CNL), Department of Physics, University of Guilan, P. O. Box 41335-1914, Rasht, Iran}
\author{Anita Yadav}
\affiliation{Abdus Salam International Centre for Theoretical Physics, Strada Costiera 11, 34151 Trieste, Italy}%
\author{Nata\v sa Stoji\' c}
\affiliation{Abdus Salam International Centre for Theoretical Physics, Strada Costiera 11, 34151 Trieste, Italy}
\author{H. Rahimpour Soleimani}
\affiliation{Computational Nanophysics Laboratory (CNL), Department of Physics, University of Guilan, P. O. Box 41335-1914, Rasht, Iran}
\author{Nadia Binggeli}
\affiliation{Abdus Salam International Centre for Theoretical Physics, Strada Costiera 11, 34151 Trieste, Italy}

\begin{abstract}

The emergence of hidden spin polarization in centrosymmetric nonmagnetic crystals due to local symmetry breaking has created new opportunities for potential spintronic applications and for enhancing our understanding of mechanisms to electrically manipulate spin-related phenomena. 
In this work, we investigate within density functional theory the properties of the hidden spin texture and spin-layer segregation in a prototype centrosymmetric dichalcogenide-monolayer material using an electric-field-based method.  This method is shown to yield a precise and robust  alternative to traditional layer-projected spin-polarization techniques for obtaining the intrinsic hidden spin textures in such materials. Moreover, it gives access at the same time to the spatial distribution within the monolayer of the individual spin-segregated states responsible for the hidden spin textures, not provided by other techniques. With this approach we determine and study the hidden spin textures of the upper valence bands of the  \(PtTe_2\) monolayer together with the spatial behavior of the probability densities and spin polarization densities of the corresponding maximally segregated spin states. This combined study enabled by the electric-field method yields new insights into the mechanisms controlling the spin-layer segregation and resulting hidden spin texture in such systems. We also discuss the symmetry rules governing the shape in the Brillouin zone of the hidden spin texture, which can be straightforwardly predicted within the present framework. \\

\end{abstract}


\maketitle   \newpage

\section{\label{sec:level1}Introduction}

Spin-orbit interaction gives rise to a variety of exciting phenomena and properties in crystalline materials, which include recently discovered topological phases of matter and electronic spin textures~\cite{ManKooNit15,SouReyFer16,Sch17}. In particular, the spin texture of band states in momentum space arising from spin-momentum locking has become a strong focus of interest lately, as it is crucial for understanding many spin-related phenomena and determines how electronic spin-polarized currents may be manipulated for spintronic devices. 

Spin-orbit coupling (SOC) can generate spin polarisation in non-magnetic crystals lacking inversion symmetry through effects such as the Rashba and Dresselhaus effects \cite{Dre55, BycRas84}, which induce a momentum-dependent splitting of the otherwise doubly spin-degenerate bands. The properties of the resulting spin textures in momentum-space have been extensively investigated for various classes of non-centrosymmetric three-dimensional (3D) and two-dimensional (2D) crystalline systems~\cite{Sch17,LvMufQia15,TaoPauTul17,RinVarAsa18,TaoTsy18,SakHirTak20,YanCanLuc21,AcoYuaDal21,BusMadGoh23,FraFarZol24}. Symmetry-based rules have also been established as guidelines for the shapes of the spin textures of non-centrosymmetic crystals in their Brillouin zone (BZ) \cite{AcoYuaDal21}.

The recently discovered hidden spin polarization in inversion-symmetric non-magnetic crystals has demonstrated that SOC-generated spin polarization can also arise due to specific local atomic-site asymmetry in crystals that maintain global inversion symmetry~\cite{ZhaLiuLuo14, YuaLiuZha19}.  
Such spin polarization is hidden in momentum-space, as bands remain doubly degenerate with zero net spin resulting from the pair of degenerate states with opposite spins, but is non-vanishing in real space. This real-space spin polarization stems from the existence of spatial segregation induced by SOC of the spin-state wavefunctions in inversion symmetric sectors of the crystal \cite{ZhaLiuLuo14, YuaLiuZha19}.  In the case of 2D materials, this leads to an interesting spin-layer locking effect \cite{YaoWanHua17}. 
The hidden spin polarization and spin-state segregation are remarkable phenomena that can open new possibilities for applications  in  spintronics  and  enhance our  understanding of  mechanisms to electrically manipulate  spin-related phenomena~\cite{LiuZhaJin15, ZhaLiuSun20, ZutFabSar04, KookwoEom09, SouReyFer16, LinYanZha19, LiuZhaZun15,GuaXioWan22}. 

Experimental and theoretical density function theory (DFT) studies have reported hidden spin polarisation~\cite{ZhaLiuLuo14, YuaLiuZha19,SanCasLan16} and spin textures in a number of inversion-symmetric materials, notably van der Waals layered materials and 2D materials \cite{RilMazDen14,GehAguBih16,YaoWanHua17,CheSunChe18,LinWanXu20,ZhaZhaHao21,LeeKimKwo23}, and have advanced our understanding of the related spin polarisation physics. 
However, in general, hidden spin textures and the spatial behavior of the related spin states are less accessible and less well known than their counterparts in non-centrosymmetric crystals, which may also hinder a detailed microscopic understanding of mechanisms involved in determining   their properties. 

Usually, spin textures can be measured experimentally for occupied states by means of 
 spin-angle-resolved photoemission spectroscopy (SARPES), and this technique has been widely used to determine the spin textures of a large range of non-centrosymmetric crystals~\cite{LvMufQia15,RinVarAsa18,SakHirTak20,BusMadGoh23}.
 Hidden spin textures of centrosymmetric crystals are more difficult in general to obtain.
In the case of centrosymmetric crystals, hidden spin texture and spin-layer locking could be evidenced by SARPES in several 2D materials and layered materials~\cite{RilMazDen14,GehAguBih16,YaoWanHua17,ZhaZhaHao21}, exploiting the fact that penetration depth of the photoemission process probes preferentially the outermost layer~\cite{ZhaZhaHao21}. In particular, Yao et al. have reported  experimental observation of  helical spin texture and spin-layer locking in centrosymmetric monolayer $PtSe_2$~\cite{YaoWanHua17}. 
Theoretically, the hidden spin texture in inversion-symmetric crystals is also not as well defined as in the standard case of non-centrosymmetric crystals and cannot be simply evaluated as in the latter case.  This is because separately the spin polarization vectors of the spin-degenerate states are arbitrary, as they depend on the choice of the two orthogonal eigenstates in the degenerate subspace (while their sum is invariant, but zero).  In DFT studies of hidden spin polarization of 2D systems and layered materials, the spin textures for such spin-degenerate bands have been generally evaluated thus as the total spin polarisation in real space of the degenerate-band states projected on (integrated in) each of the two inversion-symmetric layers (sectors) of the crystal~\cite{RilMazDen14,YaoWanHua17,CheSunChe18,LinWanXu20,LeeKimKwo23}. 

In the present work, we investigate within DFT the hidden spin texture and spin-layer segregation behavior for the prototype centrosymmetric monolayer  \(PtTe_2\) using a different approach. We apply a tiny electric field normal to the monolayer, which is shown to be an effective, robust approach for determining the hidden spin texture of the pristine monolayer and at the same time giving access to the spatial distribution within the monolayer of the segregated spin states causing the hidden spin polarisation. With this approach, we provide the hidden spin textures of the upper valence bands of the  \(PtTe_2\) monolayer and examine the corresponding spatial behavior of the probability densities and spin polarization density of the segregated spin states at different points of the BZ.  This yields new insights into the mechanisms determining the segregation behavior and resulting properties of the hidden spin polarization texture. We also discuss the shape in the BZ of the hidden spin texture, which can be straightforwardly understood within the present approach. 

\section{\label{sec:level2}Computational details}

Our first-principle calculations are performed within the framework of DFT as implemented in the Quantum ESPRESSO package \cite{GiaBarBon09}, with the plane-wave basis set. The generalized gradient approximation (GGA) with Perdew-Burke-Ernzerhof functional is used to compute exchange-correlation energy of the electrons \cite{PerBurErn96}.   SOC was incorporated using fully-relativistic projector-augmented-wave (PAW) pseudopotentials. The plane-wave kinetic energy cutoffs for the wavefunctions and charge density are set to 75~Ry and 450~Ry, respectively. BZ sampling is done using a $\Gamma$-centered Monkhorst–Pack $k$-grid of 10 × 10 × 1. A vacuum of 25~\AA\ thickness is used to eliminate interactions between adjacent perpendicular slabs. In the structural relaxations, all coordinates are relaxed until the magnitude of the forces acting on each atom is less than 0.1~mRy/Bohr. 


\begin{figure*}  
    \centering
    \includegraphics[width=0.8\linewidth]{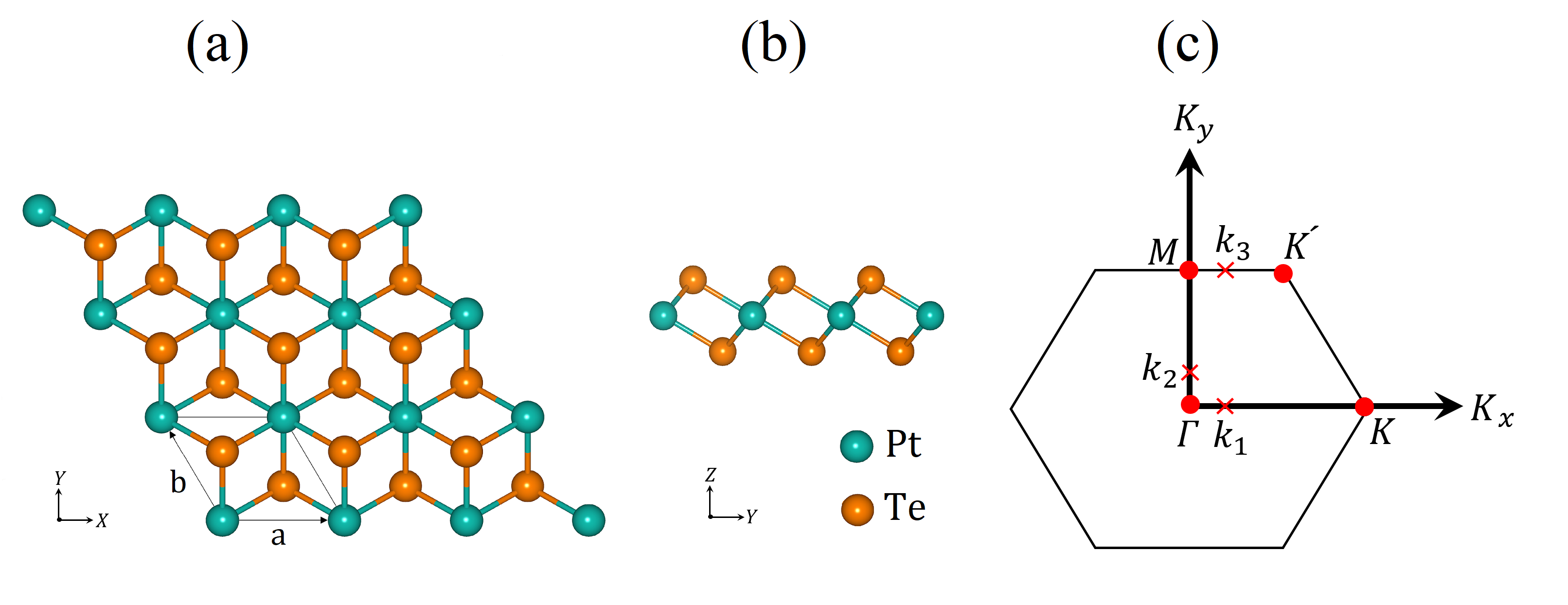}
    \caption{(a) Top and (b) side view of the $1T$-$PtTe_2$ monolayer structure, with lattice vectors \textbf{a} and \textbf{b} indicated in the figure. (c) The first Brillouin zone of the monolayer $PtTe_2$ , with indicated in red the $k$-points considered in this study to investigate the spatial behavior of the segregated spin states. These points include the four high symmetry points  marked with red circles: $\Gamma$, K, M and K', and the three $k$-points, $k_1$, $k_2$, and $k_3$, marked with red crosses. }
    \label{structure}
\end{figure*}


Fig.~\ref{structure} shows (a) the top view and (b) the side view of the optimized atomic structure of  \(PtTe_2\) monolayer. The equilibrium lattice constant from our DFT calculations is 4.03~\AA, which is consistent with previous calculations for the monolayer and in agreement with the experimental value for bulk \(PtTe_2\) \cite{VilCriHua19}. 

The spin texture of the bands in $k$-space is determined by computing the expectation values of the spin operators 
($\frac{1}{2} \hat{\sigma}_{\alpha}$, with $\alpha = x, y, z$), in the Bloch spinor eigenfunctions $\psi_{n,\bm{k}}(\bm{r})$: 
\begin{equation}
S_{\alpha}(n,\bm{k})=\frac{1}{2}\int\limits_{\Omega}^{ } \psi^{+}_{n,\bm{k}}(\bm{r})\hat{\sigma}_{\alpha}\psi_{n,\bm{k}}(\bm{r}) d{\bm r}, 
\end{equation} 
where the integration is over the supercell $\Omega$; $n$ denotes the band index, and $\hat{\sigma}_{\alpha}$ are the Pauli matrices. The  Bloch spinor states $\Psi_{n, {\bm k}}({\bm r})$ are assumed to be normalized over the supercell. The spin polarization vectors $\bm{S}(n,\bm{k})$ of the spin texture are evaluated on a grid of ${\bm k}$ points in the BZ. 

For a complete understanding of the spin-polarization behavior, we also examine for specific states $\psi_{n,\bm{k}}(\bm{r})$ 
the probability density and local magnetization density vector, whose  components  are determined as: 
\begin{equation}
m_{\alpha}^{(n,{\bm k})}(\bm{r})=\mu_{B}\psi^{+}_{n,\bm{k}}(\bm{r})\hat{\sigma}_{\alpha}\psi_{n,\bm{k}}(\bm{r}), 
\end{equation}
 where $\mu_{B}$ represents the Bohr magneton. 

The probability density and local magnetization density components are investigated at selected $k$ points of the BZ of the monolayer, indicated in Fig.~\ref{structure} (c). These points include the high symmetry points of the BZ: $\Gamma$ (0, 0, 0), $K$ (0.667, 0, 0), $M$ (0, 0.577, 0), and $K'$ (0.333, 0.577, 0), as well as specific points $k_{1}$ (0.167, 0, 0), $k_{2}$ (0, 0.144, 0), and $k_{3}$ (0.167, 0.577, 0) located along the high-symmetry lines $\Gamma$- $K$,  $\Gamma$-$M$, and $M$-$K'$, respectively. All coordinates are given in units of $2\pi/a$. 


\begin{figure}[htb]
    \centering
    \includegraphics[width=0.45\textwidth]{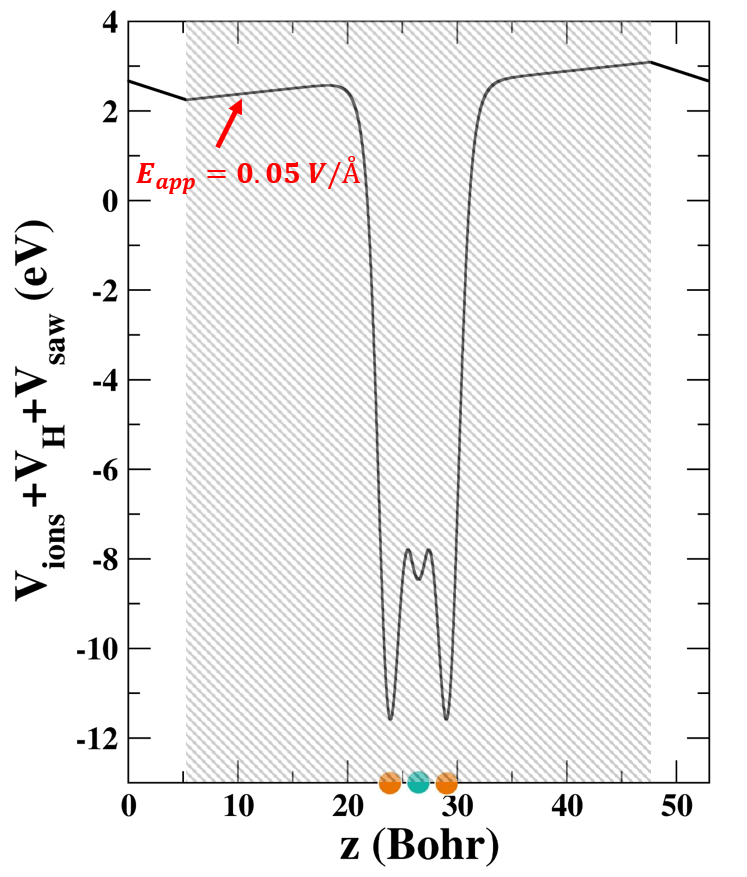}
    \caption{Planar averaged plots of the total electrostatic potential along the $\hat{z}$ direction, including the local component of the ionic pseudopotential ($V_{ions}$), Hartree potential ($V_{H}$), and sawtooth potential ($V_{saw}$) for an applied electric field, $E_{app}=$~0.05 V/{\AA}. The  gray hatched region indicates the part  subjected to the electric field. The orange and cyan circles show the positions of the Te and Pt atoms, respectively, along the  $\hat{z}$ direction. The  red arrow represents the slope corresponding to the electric field.  }
    \label{V_ion+H+saw}
\end{figure}


The electric field is modeled by a sawtooth potential in the direction perpendicular to the monolayer plane (the $\hat{z}$ direction). Fig.~\ref{V_ion+H+saw} shows the planar average of the electrostatic potential along the $\hat{z}$ direction, represented as $V_{ions} + V_{H} + V_{saw}$. Here, $V_{ions}$ represents the local part of the ionic pseudopotential, $V_H$ is the Hartree potential, and $V_{saw}$ is the sawtooth potential due to the applied external field $E_{app}$. 
In our calculations, the external field $E_{app}$ is applied within the hatched region indicated in the figure.  A dipole correction is included  in the vacuum region, compensating for the monolayer polarization dipole induced by the electric field, in order to keep the slope of the electrostatic potential in the hatched region at the constant $E_{app}$ value. 

\section{\label{sec:level3}Results and discussion}

To investigate the hidden spin texture and the spatial segregation of the states that give rise to the spin-layer locking effect and to the hidden spin texture, we have applied a small electric field $E_{app}=$~0.05 V/{\AA} perpendicular to the monolayer. In order to gauge the electric field and validate the intrinsic character of the segregated states and of their spatial distribution and spin textures, we considered different values of the electric field ranging from 0.05 V/{\AA} to 0.2 V/{\AA}. 

\begin{figure}[htb]
    \centering
    \includegraphics[width=0.53\textwidth]{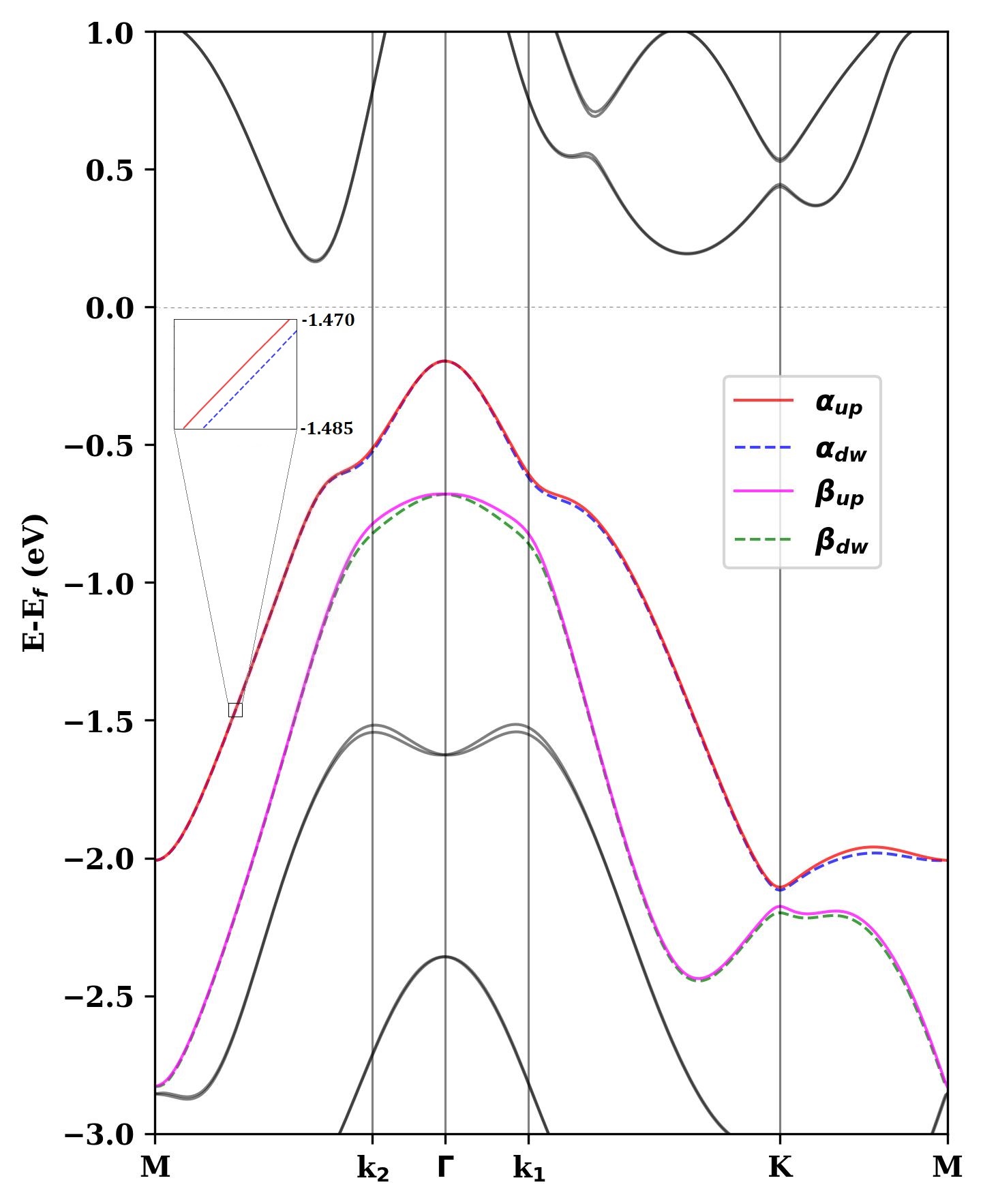}
    \caption{ The band structure of  $PtTe_2$  monolayer with SOC under  an electric field of 0.2 V/Å. The first and second upper valence bands, indicated by a solid red and a dashed blue line, are labeled as $\alpha_{up}$ and $\alpha_{dw}$, respectively. 
    The third and fourth upper valence bands, shown by a solid purple  and green dashed line, are denoted as  $\beta_{up}$ and $\beta_{dw}$, respectively. The inset shows a magnified view of the bands within a smaller energy range, outlined by the solid black box, highlighting the splitting of the $\alpha_{up}$ and $\alpha_{dw}$ bands.}
    \label{BANDS}
\end{figure}

Fig. \ref{BANDS} shows the band structure of the monolayer in the presence of an applied electric field  $E_{app}=$~0.2 V/{\AA}.
 Before the application of the electric field, the presence of time-reversal  symmetry
 and inversion symmetry
 in the system results in each band within the band structure having a twofold degeneracy. 
Upon application of the electric field, inversion symmetry is lost, resulting in the disappearance of the double degeneracy of the bands. The corresponding splitting of the bands increases in direct proportion to the magnitude of the applied electric field, for the fields considered. 
In our study, we focus on the first and second upper valence bands (at $E_{app}=$~0) of the \(PtTe_2\) monolayer, which are referred to as the $\alpha$ and $\beta$ bands, respectively. These are the bands closest to the bandgap displaying largest spin-layer locking. In the presence of the electric field the $\alpha$ and $\beta$   bands  split into $\alpha_{up}$,  $\alpha_{dw}$  bands and  $\beta_{up}$, $\beta_{dw}$ bands, respectively, indicated by the color lines in Fig.~\ref{BANDS}, where the label {\it up} ({\it dw}) indicates the upper- (lower-) energy split band.
The splitting is $k$ dependent and, although small, is present everywhere in the BZ, except at the time-reversal-invariant points 
 $M$ and $\Gamma$. The splitting  between the $\beta$ bands  tends to be greater in general than that of the $\alpha$  bands, as can be noticed  in Fig.~\ref{BANDS}, especially near the  $k_{1}$ and $k_{2}$ points (marked in Figs.~\ref{structure}(c)), which are located respectively along the $\Gamma$--$K$ and $\Gamma$--$M$ lines, 
and at the $K$ point.

\begin{figure*}
    \centering
    \includegraphics[width=0.7\textwidth]{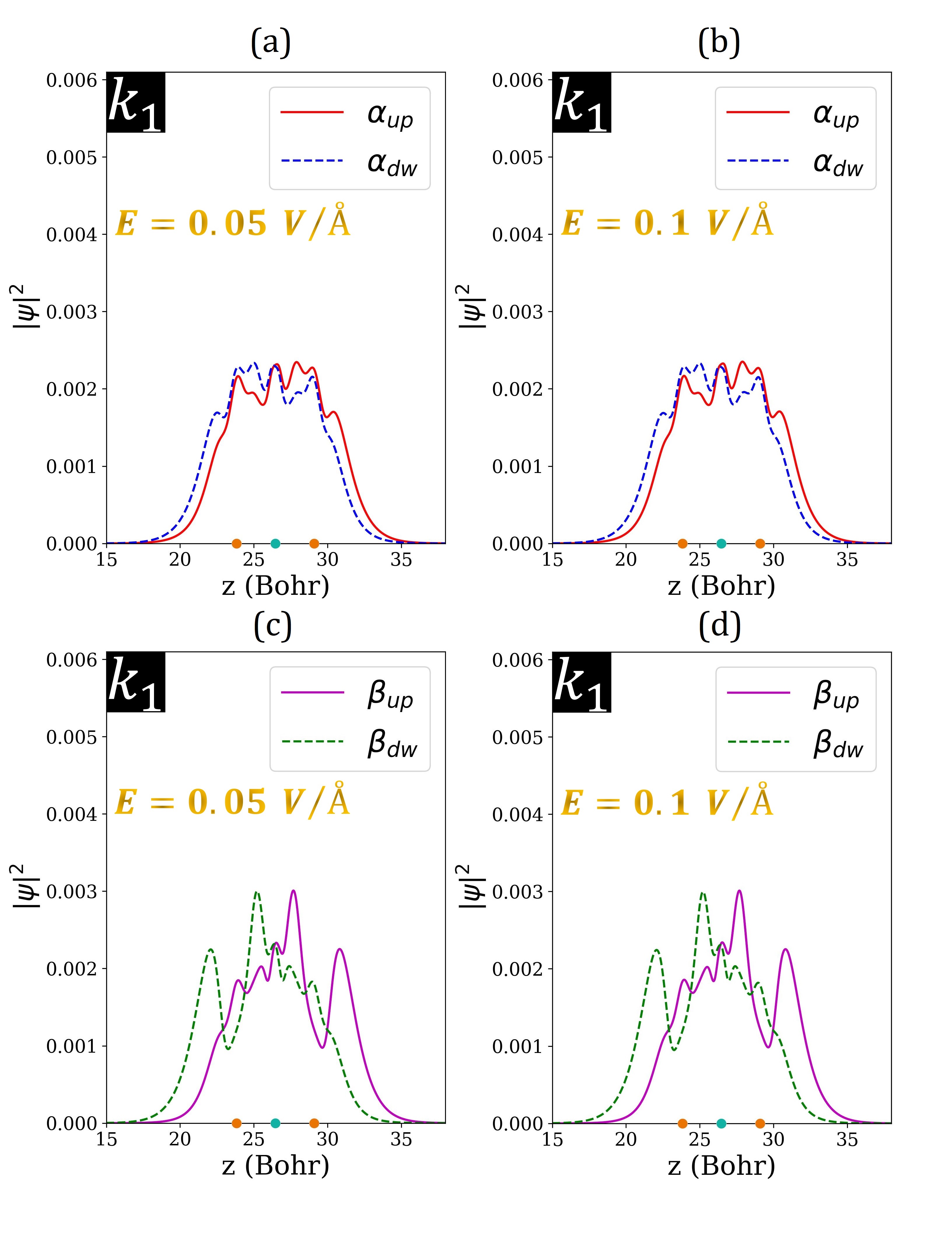}
    \caption{Planar average of the probability density (in ${Bohr}^{-3}$)  for the electronic states of the $\alpha$  (panels (a)-(b)) and $\beta$ (panels (c)-(d)) valence bands of the $PtTe_2$ monolayer at the point $k_{1}~(0.167,0,0)2\pi/a$ in the presence of different electric fields, (a) and (c) $E_{app}=$~0.05 V/{\AA}, (b) and (d)  $E_{app}=$~0.1 V/{\AA}. The orange and cyan circles show the positions of the Te and Pt atoms, respectively, along the z direction. }
    \label{prob_k1}
\end{figure*}

In Figs.~\ref {prob_k1} (a) and (c), we show the planar
average of the probability density \ensuremath{|\psi|^2} 
obtained for the split states of the $\alpha$ and $\beta$ bands at the ${\bm k}$ point $k_{1}$ (along the $\Gamma - K$ line) 
in the presence of the electric field $E_{app}=$~0.05 V/{\AA}. 
The probability distribution of the split states indicates  a segregation, within the $PtTe_{2}$ monolayer, of the $\alpha_{up}$ and $\alpha_{dw}$ ($\beta_{up}$ and $\beta_{dw}$) states towards the top and bottom Te layers, respectively. This segregation is more pronounced in the case of the $\beta$ band than for the $\alpha$ band. In Figs.~\ref {prob_k1} (b) and (d), we display the planar-averaged probability distributions of the states of the  $\alpha$ and $\beta$ bands calculated at the same $\bm{k}$ point, but with the magnitude of the applied field doubled ($E_{app}=$~0.1 V/{\AA}). The results show virtually no change in the distribution of the segregated states. 
In fact, we find that up to an amplitude of about 0.2 V/{\AA} of the applied field, the two distributions remain nearly identical to the initial segregated distributions obtained with the smallest electric field. This indicates that the result on the probability segregation in the smallest fields, in Figs.~\ref {prob_k1}, corresponds to the limit  $E_{app} \rightarrow 0 $ and that the segregated states are intrinsic to the pristine $PtTe_2$ monolayer. We also examined the effect of the electric fields $E_{app}=$~0.05 V/{\AA} and $E_{app}=$~0.1 V/{\AA} on the probability distributions of other eigenstates of the $\alpha$ and $\beta$ bands at different $\bm{k}$ points marked in Fig.~\ref{structure}(c) and similarly found no noticeable influence of the amplitude of the finite applied fields on the probability distributions of the segregated states.  

Hence the effect of the electric field, which slightly shifts to higher energy the potential of the top relative to the bottom Te layer, is simply to pick out from the subspace of the degenerate eigenstates of the unperturbed system (at a given $\bm{k}$  point) the pair of spin states with maximal segregation on the two layers. When no field is applied, instead, the eigenstates are arbitrary linear combinations of the maximally segregated states. They have probability distributions that are intermediate between, on one hand, those of two identical fully symmetric distributions [see, e.g., Figs. S3 (b) and (g) of the Supplementary Material (SM), as examples of results obtained without field] corresponding to the average between the distributions of the two segregated states, and, on the other hand, the two separated distributions of the maximally segregated states [Figs.~\ref {prob_k1} (a) and (c)]. We note that the electric-field method can be expected to yield accurately the intrinsic (i.e., zero-order in the field) maximally segregated states of the pristine system, as long as the band splitting is linear (of first order) in the electric field ---given the latter implies that the first-order correction to the segregated electronic states has a negligible impact.

\begin{figure*}
    \centering
    \includegraphics[width=1\linewidth]{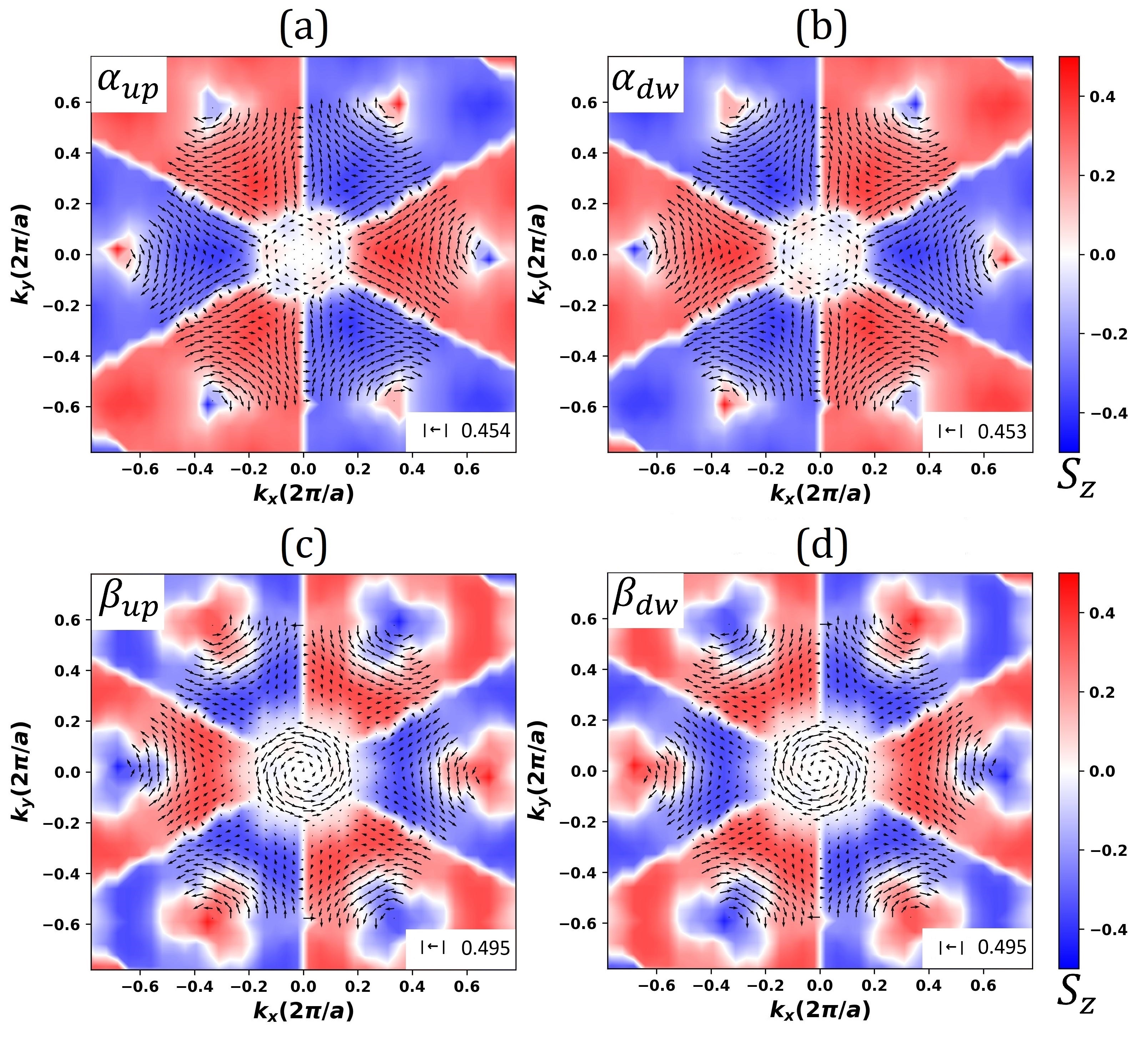}
    \caption{Spin texture of the $\alpha$ and $\beta$  valence bands of the $PtTe_2$ monolayer in the presence of a small electric field ($E_{app}=$~0.05 V/{\AA}). The in-plane spin components $(S_{x}, S_{y})$ are shown by black arrows at the $k$-grid points in the 2D BZ, while the out-of-plane spin texture component is represented by the map of  $S_{z}$ isovalues. The length of the arrows represents the magnitude of the in-plane spin components, with the scale provided in the bottom-right corner of the plot.}
    \label{spintexture_E0.05}
\end{figure*}

Having seen the robustness of the inherent segregated states with the fields considered,  we evaluated the hidden spin textures of the $\alpha$ and $\beta$ bands using the electric field $E_{app}=$~0.05 V/{\AA}. The resulting spin textures for the $\alpha_{up}$, $\alpha_{dw}$,   $\beta_{up}$, and $\beta_{dw}$ states of the $PtTe_2$ monolayer are presented in Fig.~\ref{spintexture_E0.05}(a), (b), (c), and (d), respectively. Doubling the electric field yielded no noticeable change in the spin textures. Without
 field, instead, the spin textures have  arbitrary distributions, see, e.g.,  Fig. S1 in the SM.


After applying the electric field, the spin textures of the
$\alpha$ and $\beta$ bands, in Fig.~\ref{spintexture_E0.05}, have properties and trends qualitatively consistent with those observed in the hidden spin textures of other transition-metal  dichalcogenides obtained using the layer-projected spin polarization approach \cite{RilMazDen14,YaoWanHua17,CheSunChe18}.
In particular, around the $\Gamma$ point, the in-plane spin texture is very small (almost vanishing) for the $\alpha$ states, while the $\beta_{up}$ and $\beta_{dw}$ states (segregated on the top and bottom Te layer, respectively) display hefty helical spin textures with tangential in-plane spin polarization vectors rotating  
 counterclockwise ($\beta_{up}$) and clockwise ($\beta_{dw}$).
This is consistent with the hidden Rashba-type spin textures produced around $\Gamma$ by the local Rashba effect \cite{ZhaLiuLuo14}, caused by the out-of-plane local electric field induced on each surface chalcogen layer by the presence of the other chalcogen layer and the transition-metal layer~\cite{YaoWanHua17}. Such local electric fields have the same magnitude and opposite signs on the two Te layers, leading to the observed opposite chiralities of the in-plane helical spin textures around $\Gamma$ for the top ($\beta_{up}$) and bottom ($\beta_{dw}$) Te layers. 

The extremely small in-plane spin texture near the center of the BZ for the states of the $\alpha$ band relative to the $\beta$ band is related to their specific spin-orbital characters \cite{CheSunChe18}. The states of the $\alpha$ and $\beta$ bands near $\Gamma$ are mainly composed of Te-$5p$ bonding states with total spin-orbital momenta J=3/2, $J_z = \pm$ 3/2 and  J=3/2, $J_z = \pm 1/2$, respectively, see Fig. S2 in the SM. This leads to a nearly vanishing in-plane spin texture for the $\alpha$ states (contrary to the $\beta$ states) given the rules:  $\langle J, J_{z}|\hat{\sigma}_{\alpha}|J, J'_{z}\rangle = 0$ for $\alpha = x,y$, when $\Delta J_z =J_z-J'_z \neq \pm 1$.  

The high-symmetry $K$ and $K'$ points, unlike other band-extrema points such as $\Gamma$ and $M$, lack  time-reversal symmetry (with/without $E_{app}$). Therefore, at and near the $K$, $K'$ points one observes non-magnetic Zeeman type of spin textures~\cite{YuaBahMor12,MerFazDal19}, related to the  spin splitting of the bands at and near these band extrema. The fact that the $\bm{k}$-point groups of $K$ and $K'$ include (with/without  $E_{app}$) a rotational axis (C$_3$) perpendicular to the layer implies that the corresponding spin-split segregated states $\alpha_{up}$ and $\alpha_{dw}$ ($\beta_{up}$ and $\beta_{dw}$)  with opposite spins on the two Te layers must exhibit purely out-of-plane spin-polarization components at $K$,  $K'$  \cite{AcoYuaDal21}, as observed in Fig.~\ref{spintexture_E0.05} (and further discussed below). In addition, a local Rashba effect is also observable in the in-plane spin texture at some distance around the $K$, $K'$ points for all four bands.

Within the electric-field approach, the symmetry properties of the hidden spin textures of 2D systems, controlling the general shape of their hidden spin textures in the BZ, can be straightforwardly understood. To the best of our knowledge such  symmetry properties have never been discussed so far. In fact, because the system with vertical electric field is non centrosymmetric (due to the field), the resulting spin texture must comply with the symmetry rules already established for spin textures of non-magnetic non-centrosymmetric crystals  \cite{AcoYuaDal21}, provided one considers the $\bm{k}$ point-group symmetries of the system with $E_{app} \neq 0$. 

 In particular, for non-magnetic non-centrosymmetric materials, the symmetry rules  for the spin polarization vector $\bm{S}(\bm{k})$ impose that  \cite{AcoYuaDal21}: (i) when the  $\bm{k}$ point lies within a mirror plane, $\bm{S}(\bm{k})$ should be oriented perpendicular to the mirror plane; and (ii) when the $\bm{k}$ point is on a rotational axis, $\bm{S}(\bm{k})$ must be parallel to the rotational axis. 
In our case, the lines from the $\Gamma$ to the $M$ points (to the 6 equivalent $M$ points on the sides of the BZ) belong to vertical mirror planes for the system with $E_{app}$. Therefore,  for $\bm{k}$ points on those lines, $\bm{S}(\bm{k})$ must be perpendicular to   the mirror plane. This explains the triangular array of white lines ($S_{z}=0$) seen in the spin textures of  Fig. \ref{spintexture_E0.05},  connecting the $\Gamma$ to the $M$ points along these mirror planes. In addition, the $\Gamma$, $K$, and $K'$ points themselves are located on vertical C$_3$ rotational axes of $\pm 120{\degree}$ rotation operators belonging to the point groups of these $\bm{k}$ points in presence of $E_{app}$. Apart from giving rise to vanishing in-plane spin components $(S_x, S_y)$ at the $K$, $K'$ points and around $\Gamma$, this also accounts for the observed invariance of the spin textures under $\pm 120{\degree}$ rotations about these vertical axes at $\Gamma$, $K$, and $K'$. 
Furthermore, for non-magnetic non-centrosymmetric crystals in general \cite{Ghosh_sym}, time reversal symmetry imposes  ${\bm S}(-{\bm k}) = - {\bm S}({\bm k})$, and hence the out-of-plane spin polarizations and chiralities of the helical in-plane spin texture are observed to be opposite at and around the $K$ and $K'$ points. 

Aside from providing the hidden spin textures, the electric field approach also gives access to the spatial behavior of the probability density and spin-polarization density of the individual maximally segregated states responsible for the spin-layer locking effect and hidden spin texture. This spatial behavior of the individual segregated states is not available with other methods to obtain hidden spin textures, and can provide  insights into the mechanisms controlling the segregation and hidden spin texture. 

\begin{figure*}
    \centering
    \includegraphics[width=1\linewidth]{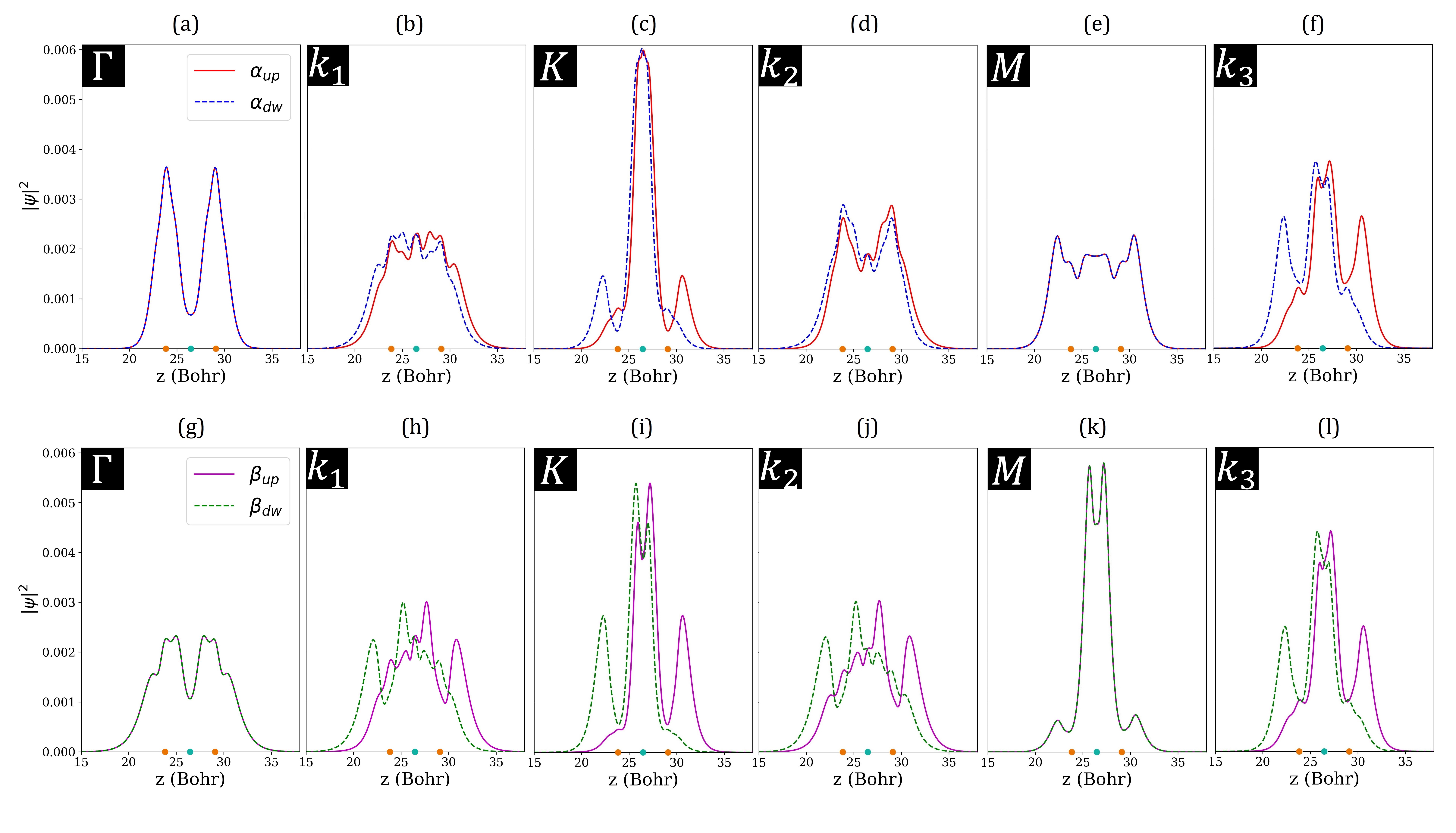}
    \caption{Planar average of the probability density (in ${Bohr}^{-3}$) for the electronic states of the $\alpha$ (panels (a)-(f)) and  $\beta$ (panels (g)-(l)) valence bands of $PtTe_2$ monolayer in the presence of the small electric field ($E_{app}=$~0.05 V/{\AA}), at different Brillouin-zone $\bm{k}$ points: (a) and (g) $\Gamma~(0,0,0)2\pi/a$, (b) and (h) $k_{1}~(0.167,0,0)2\pi/a$, (c) and (i)  $K~(0.667,0,0)2\pi/a$, (d) and (j)  $k_{2}~(0,0.144,0)2\pi/a$,(e) and (k) $M~(0,0.577,0)2\pi/a$ and (d) and (l)  $k_{3}~(0.167,0.577,0)2\pi/a$. The orange and cyan circles show the positions of the Te and Pt atoms, respectively, along the z direction. }
    \label{prob_total_E0.05}
\end{figure*}

In Fig.~\ref{prob_total_E0.05},
 we display the planar average of the probability density of the maximally segregated states (obtained with $E_{app}$=0.05 V/{\AA}) for the $\alpha$ and $\beta$ bands at selected $\bm{k}$ points of the BZ [indicated in Fig. \ref{structure} (c)]. As mentioned before, such states ---picked out from the degenerate subspace by a tiny electric field--- are essentially intrinsic states of the pristine bilayer.
 Fig. \ref{prob_total_E0.05} indicates that, apart from $\Gamma$ and $M$, characterized by two identical symmetric probability densities for the doubly degenerate states (due to time reversal $T$ and inversion $I$ symmetry at those $\bm{k}$ points at $E_{app}$=0), and hence showing no segregation, all other $\bm{k}$ points display {\it up} and {\it dw} band states having distinct probability densities, with segregation towards top and bottom Te layers, respectively. We note that when no field is applied, instead, any $\bm{k}$ point can display identical symmetric probability densities of the degenerate 
band states, see Fig. S3 of the SM.


\begin{figure*}
    \centering
    \includegraphics[width=1\linewidth]{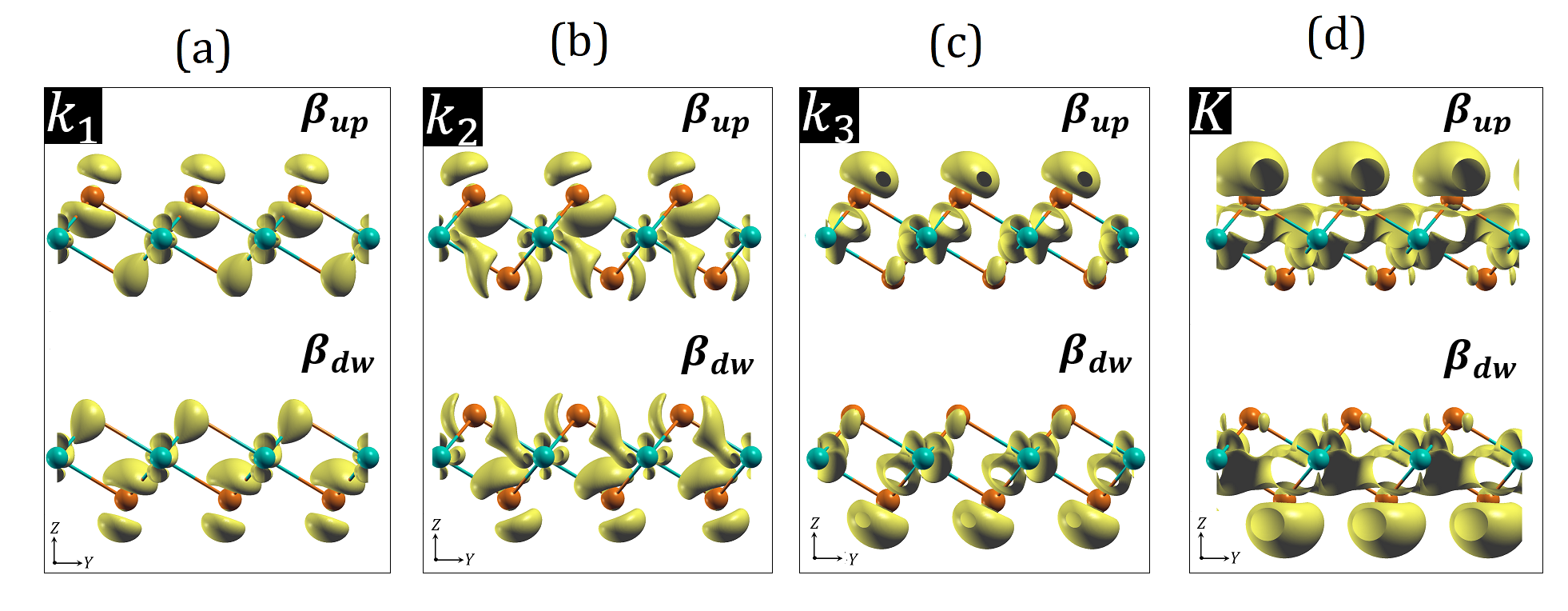}
    \caption{Iso-surface plots of the probability density of the $\beta_{up}$ (top row) and $\beta_{dw}$ (bottom row) states of the $PtTe_2$ monolayer in the presence of the small electric field ($E_{app}=$~0.05 V/{\AA})  at the points $k_{1}~(0.167,0,0)2\pi/a$, $k_{2}~(0,0.144,0)2\pi/a$, $k_{3}~(0.167,0.577,0)2\pi/a$ and $K~(0.667,0,0)2\pi/a$. The orange and cyan circles show the Te and Pt atoms, respectively. The iso-surface value is 0.005  $e/{Bohr}^3$ for (a) and (b), 0.0035 $e/{Bohr}^3$  for (c), and 0.0012 $e/{Bohr}^3$ for (d). 
    }
    \label{prob_xsf}
\end{figure*}


The asymmetry between the probability densities of the {\it up} and {\it dw} segregated states, in Fig.~\ref{prob_total_E0.05}, is observed mostly in the region of the Te orbitals, corresponding to the atomic sites with broken local inversion symmetry. 
The segregation is seen to be strongest at the $K$ and $k_3$ points for both the $\alpha$ and $\beta$ bands, and also at the $k_1$ and $k_2$ points for the $\beta$ band. This corresponds to the regions of these bands where the non-magnetic Zeeman effect and/or Rashba effects are strongest, as measured by the intensity of the respective vertical and/or helical in-plane components of the spin textures in Fig.~\ref{spintexture_E0.05}. It should be noted though that even in the cases of strongest segregation observed in Fig.~\ref{prob_total_E0.05}, the separation of the maximally segregated states is not complete, i.e., a substantial  inherent overlap remains between the planar averaged probability densities of the {\it up} and {\it dw} states. This is apparent, in particular, in the region of the Pt orbitals, which mainly contribute to the probability density at and near the zone-edge $K$ point. This is also present however in the Te regions, to various degrees which depend on the band and $\bm{k}$ point of the segregated states.

Inspection of the probability-density profile of each segregated state, in Fig.~\ref{prob_total_E0.05}, indicates that the segregation corresponds not only to an enhanced probability density on one of the Te layer at the expense of the other, but also to a qualitative change in the type of orbitals on the two Te layers in the same state. In fact, for a given state, the Te layer with overall increased probability density, in Fig.~\ref{prob_total_E0.05}, exhibits predominantly out-of-plane Te $p_z$ orbitals, as indicated by the local minimum in the probability density at the position of that atomic layer, whereas the Te layer with overall decreased probability density displays mainly in-plane $p_{\parallel}$  orbitals, as shown by the local probability-density maximum at the position of that layer. This qualitative difference in the type of $p$ orbitals on the two Te
layers in the $PtTe_2$ monolayer is further evidenced in Fig.~\ref{prob_xsf}, where the iso-surface probability density of some of the maximally segregated  states of the $\beta$ band are displayed. 

\begin{figure*}
    \centering
    \includegraphics[width=1\linewidth]{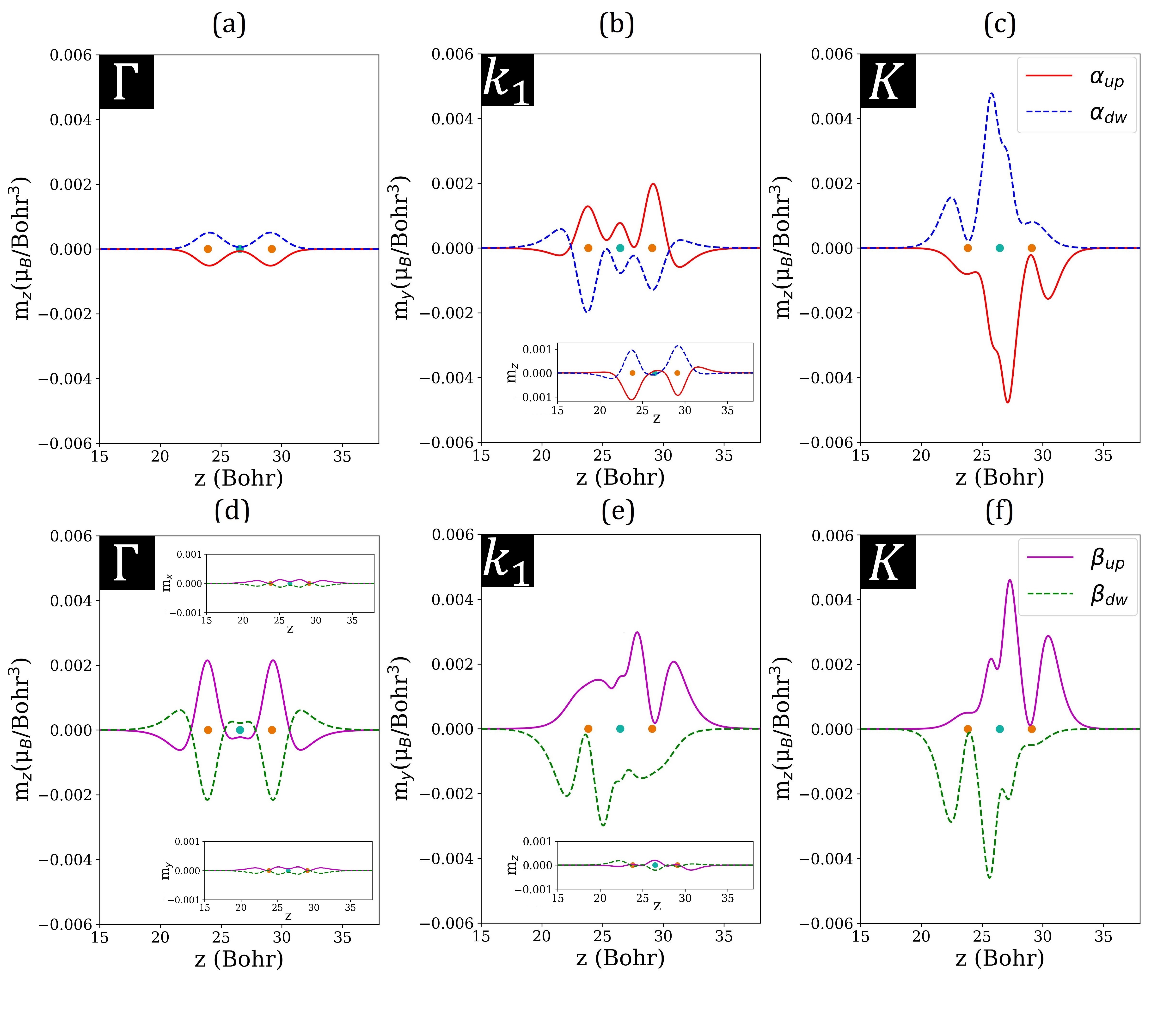}
    \caption{Planar average of the magnetization-density components for the electronic states of the $\alpha$ (panels (a)-(c)) and  $\beta$ (panels (d)-(f)) valence bands of $PtTe_2$ monolayer at the points $\Gamma$~$(0,0,0)$, $k_{1}~(0.167,0,0)2\pi/a$, $K~(0.667,0,0)2\pi/a$ in the presence of the small electric field ($E_{app}=$~0.05 V/{\AA}). The orange and cyan circles show the positions of the Te and Pt atoms, respectively, along the z direction. }
    \label{mag_G_k1_K}
\end{figure*}

\begin{figure*}
    \centering
    \includegraphics[width=1\linewidth]{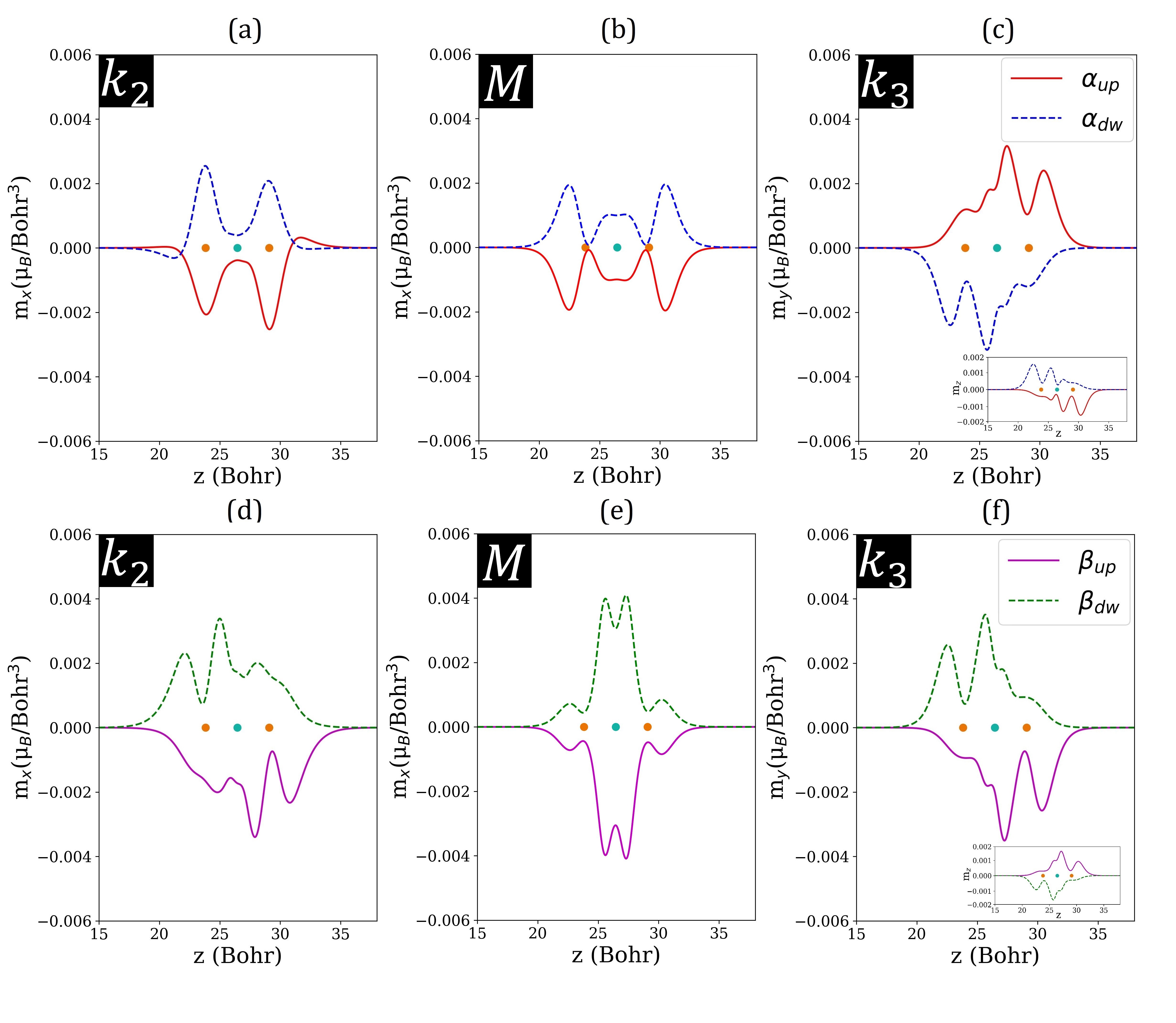}
    \caption{Planar average of the magnetization-density components for the electronic states of the $\alpha$ (panels (a)-(c)) and  $\beta$ (panels (d)-(f)) valence bands of $PtTe_2$ monolayer at the points $k_{2}~(0,0.144,0)2\pi/a$, $M~(0,0.577,0)2\pi/a$ and $k_{3}~(0.167,0.577,0)2\pi/a$ in the presence of the small electric field ($E_{app}=$~0.05 V/{\AA}). The orange and cyan circles show the positions of the Te and Pt atoms, respectively, along the z direction. }
    \label{mag_k2_M_k3}
\end{figure*}

For each of the states with probability density reported in Fig.~\ref{prob_total_E0.05}, we show in Figs.~\ref{mag_G_k1_K}-\ref{mag_k2_M_k3} the planar-averaged non-zero components of the corresponding magnetization density ${\bm m}_{\bm k}({\bm r})$.   The largest component is shown in the main figure, while smaller components (when present) are given in the insets. The magnetization densities of the {\it up} and {\it dw} states are inversion-antisymmetric partners:  ${\bm m}_{up}(-{\bm r}) = - {\bm m}_{dw}({\bm r})$ at all $\bm{k}$ points, due to the $ T \cdot I$ symmetry at no field. The integration of the planar-averaged magnetization density $\bar{{\bm m}}_{\bm k}(z)$,  in  Figs. \ref{mag_G_k1_K}-\ref{mag_k2_M_k3}, 
yields for each of the band states, $\alpha_{up/dw}$, $\beta_{up/dw}$, the polarization vector ${\bm S}({\bm k})$ at that $\bm{k}$ point in the hidden spin texture of Fig.~\ref{spintexture_E0.05}, as: 
${\bm S}({\bm k}) = \frac{A}{2 \mu_B} \int \bar{{\bm m}}_{ {\bm k}}(z) dz$, where $A$ is the surface area of the monolayer unit cell. 



For the maximally segregated states in Figs. \ref{mag_G_k1_K}-\ref{mag_k2_M_k3}, one observes that, except at $\Gamma$ and for $\alpha$ states very close to $\Gamma$, the profile of the magnetization-density components tends to be single signed and similar to that of the probability density in Fig.~\ref{prob_total_E0.05}. This together with the fact that the amplitude of ${\bm S}({\bm k})$ for such states with quasi single-signed magnetization-density components are very close to 1/2 in Fig.~\ref{spintexture_E0.05} indicates that these states are nearly pure spin states. 
This is unlike what is observed without field (SM Figs. S4-S8 ), in which case the magnetization-density profiles of the states of these bands at the same $\bm{k}$ points tend to  oscillate in sign and differ qualitatively from the profiles of the probability density, see Figs. S5-S8 in the SM.  

At $\Gamma$ and $M$, because of $T$ and $I$ symmetry, the magnetization densities of the {\it up} and {\it dw} states are found to be symmetric and opposite in Figs.~\ref{mag_G_k1_K}-\ref{mag_k2_M_k3}. Consequently, there is no hidden spin polarization at those $\bm{k}$ points. At  all other $\bm{k}$ points, instead, the spatial segregation of the two spin states is clearly visible in Figs.~\ref{mag_G_k1_K}-\ref{mag_k2_M_k3}, and gives rise to the hidden spin polarization. The trend in the degree of segregation of the magnetization densities at the different $\bm{k}$ points follows that of the probability densities, for the maximally segregated states. We note that, in  line with the significant overlaps seen between the {\it up} and {\it dw} probability densities (Fig.~\ref{prob_total_E0.05}), the magnetization densities of these spin states also largely cancel in the same spatial regions (Figs.~\ref{mag_G_k1_K}-\ref{mag_k2_M_k3}).  Furthermore, for each segregated spin state,  the difference in the main type of $p$ orbital on the two Te layers appears also  clearly in the magnetization-density profiles of Figs.~\ref{mag_G_k1_K}-\ref{mag_k2_M_k3}, displaying predominantly out-of-plane Te-$p_z$ orbitals on the layer with increased magnetization density and mainly in-plane Te-$p_{\parallel}$ orbitals on the Te layer with decreased magnetization  density.

The observation that the spin-state segregation, in Figs.~\ref{mag_G_k1_K}-\ref{mag_k2_M_k3}, is only partial and of varying strength points to the presence of at least two competing mechanisms controlling the degree of segregation and hidden spin texture.  As discussed below (and further detailed in Section S4 of the SM), the  behavior of the probability-density and magnetization-density profiles, in  Figs.~\ref{prob_total_E0.05}-\ref{mag_k2_M_k3}, can be understood as the result of the competition between the formation of bonding states, induced by the inversion symmetric potential of the $PtTe_2$ monolayer, and the Rashba and non-magnetic Zeeman effects, induced by the opposite local electric fields at the atomic sites of the two Te layers. 

This is best illustrated by the spatial distributions, in Figs.~\ref{prob_total_E0.05}-\ref{mag_k2_M_k3}, of the nearly-pure maximally-segregated spin states. These states may be viewed as the remainders of the bonding states after the influence of the Rashba and non-magnetic Zeeman effects. Their profiles reveal the conflicting influence of the two types of mechanisms ~\cite{Ghosh_FM_bilayer}. On one hand, the bonding mechanism requires the same spin components on both layers in each of the spin-degenerate states and favors therefore a symmetric distribution for each of the spin states. On the other hand, the Rashba and non-magnetic Zeeman effects demand instead two opposite spin states on the two layers (because of the opposite local electric fields) with antiparallel spin polarization vectors of maximal possible amplitude along a specific direction of space, leading to spin-state segregation.  When competing, these two types of antagonistic  mechanisms are expected to lead to distributions intermediate between the above two limits, which is what one observes in the probability-density and magnetization density profiles of the segregated states in Figs.~\ref{mag_G_k1_K}-\ref{mag_k2_M_k3}.
Furthermore, as also shown in the SM (Section S4), the change in the type of Te $p$ orbitals (from $p_{\parallel}$ to $p_z$) observed between the two layers in the segregated states around $\Gamma$  can be similarly understood as a consequence of the coupling induced by the Rashba effect among states of the $\alpha$ and $\beta$ bands.

\section{\label{sec:level4}Conclusion}

We have studied the use of an electric-field-based method to calculate the hidden spin texture from first principles in a prototype centrosymmetric relativistic non-magnetic 2D material. This technique allows not only for a precise and effective determination of the intrinsic hidden spin texture, but gives access at the same time to the spatial behavior of the individual spin-segregated electronic states responsible for the hidden spin texture, not available with other techniques providing hidden spin textures. An additional advantage of this method is its ease of application and, in particular, the fact that it can be employed also with {\it ab initio} simulation packages with no implementation of hidden-spin-texture computation (via layer-projected spin-polarization of the degenerate bands). 

With this method we have determined and investigated the properties of the hidden spin texture of the centrosymmetric $PtTe_2$ dichalcogenide monolayer together with the spatial behavior of the probability densities and spin polarization density of the associated maximally segregated spin states. This combined study enabled by the electric-field method provided new insights into the mechanisms controlling the spin-state segregation and the resulting hidden spin texture and spin-layer locking behavior in such 2D materials. In addition, we showed that the symmetry properties of the hidden spin texture, controlling the shape of the hidden texture in the BZ, could be straightforwardly predicted with the present approach.

\section*{Acknowledgments}
A. D. acknowledges support from the ICTP/IAEA Sandwich Training Educational Programme (STEP). 

\section*{Data availability}
Data will be made available on request.

\bibliography{main}

\end{document}


\title{
Supplemental Material - Investigation of the intrinsic hidden spin texture and spin-state segregation in centrosymmetric  monolayer dichalcogenide: effectiveness of the electric-field approach\\
} 
\author{Ameneh Deljouifar}
 \affiliation{Computational Nanophysics Laboratory (CNL), Department of Physics, University of Guilan, P. O. Box 41335-1914, Rasht, Iran}%

\author{Anita Yadav}
 \affiliation{Abdus Salam International Centre for Theoretical Physics, ICTP, Strada Costiera 11, 34151 Trieste, Italy}%
 
\author{Nata\v sa Stoji\' c}%
\affiliation{Abdus Salam International Centre for Theoretical Physics, ICTP, Strada Costiera 11, 34151 Trieste, Italy}%

\author{H. Rahimpour Soleimani}
 \affiliation{Computational Nanophysics Laboratory (CNL), Department of Physics, University of Guilan, P. O. Box 41335-1914, Rasht, Iran}%

\author{Nadia Binggeli}
\affiliation{Abdus Salam International Centre for Theoretical Physics, ICTP, Strada Costiera 11, 34151 Trieste, Italy}%


\maketitle

\renewcommand{\thesection}{S\arabic{section}} 
\newpage
 

\section{Examples of spin textures obtained without electric field}   
As mentioned in the main text, in the absence of electric field the individual spin polarization vectors, ${\bm S}_{n}(\textbf{k})=\frac{1}{2}\int\limits_{\Omega}^{ } \psi^{+}_{n,\textbf{k}}(\textbf{r})\hat{\boldsymbol\sigma} \psi_{n,\textbf{k}}(\textbf{r}) d{\bm r}$, are arbitrary in a centrosymmetric non-magnetic system. In fact, due to the double degeneracy at any {\bf k} point of the band eigenstates (caused by the $T \cdot I$ symmetry), the individual spin polarization vectors depend on the arbitrary choice of the pair of orthogonal eigenstates in the degenerate subspace. Consequently, the resulting spin texture will also be arbitrary and depend on the particular electronic structure code and related diagonalization details. As examples, in Fig. \ref{spintexture_without-E}, we show the spin textures obtained with Quantum Espresso (QE) for the $\alpha$ and $\beta$ bands of the monolayer $PtTe_2$  in the absence of an external electric field.  
  As can be expected, these spin textures bear no resemblance with the actual hidden spin textures of the  $PtTe_2$  monolayer determined in the main text using the electric-field method. Their properties are also qualitatively different from those observed in the hidden spin textures of other monolayers of isomorphic transition-metal dichalcogenides evaluated with the layer-projected spin polarization approach \cite{RilMazDen14,YaoWanHua17,CheSunChe18}.

\section{Projected band structure}    
   Similar to most transition metal dichalcogenides, the dominant orbital contributions to the valence bands of the $PtTe_2$ monolayer originate from the outermost $d$ orbitals of the transition metal (Pt-$5d$) and the outermost $p$ orbitals of the chalcogen atom (Te-$5p$). 
  In Fig.~\ref {projected_band}, we have plotted the Te-$5p$-orbital projected band structure of the $PtTe_2$ monolayer, decomposed into J=1/2, $J_z = \pm 1/2$ (Fig.~\ref {projected_band} (a)), J=3/2, $J_z = \pm 3/2$ (Fig.~\ref {projected_band} (b)), and J=3/2, $J_z = \pm 1/2$ (Fig.~\ref {projected_band} (c)) contributions. As evident from these plots, the dominant contribution to the two highest valence bands around $\Gamma$  comes from Te-5p orbitals with J=3/2. There is also a significant difference between the upper ($\alpha$) and second-upper ($\beta$) valence bands. In the $\alpha$ band, the dominant contribution near $\Gamma$ originates from the Te-5p orbitals with J=3/2, $J_z = \pm 3/2$ , whereas in the $\beta$ band, it comes from the Te-5p orbitals with J=3/2, $J_z = \pm 1/2$.  

\section{Examples of eigenstate probability- and magnetization- density distributions without electric field}    

\begin{figure*}[h]
    \includegraphics[width=1\linewidth]{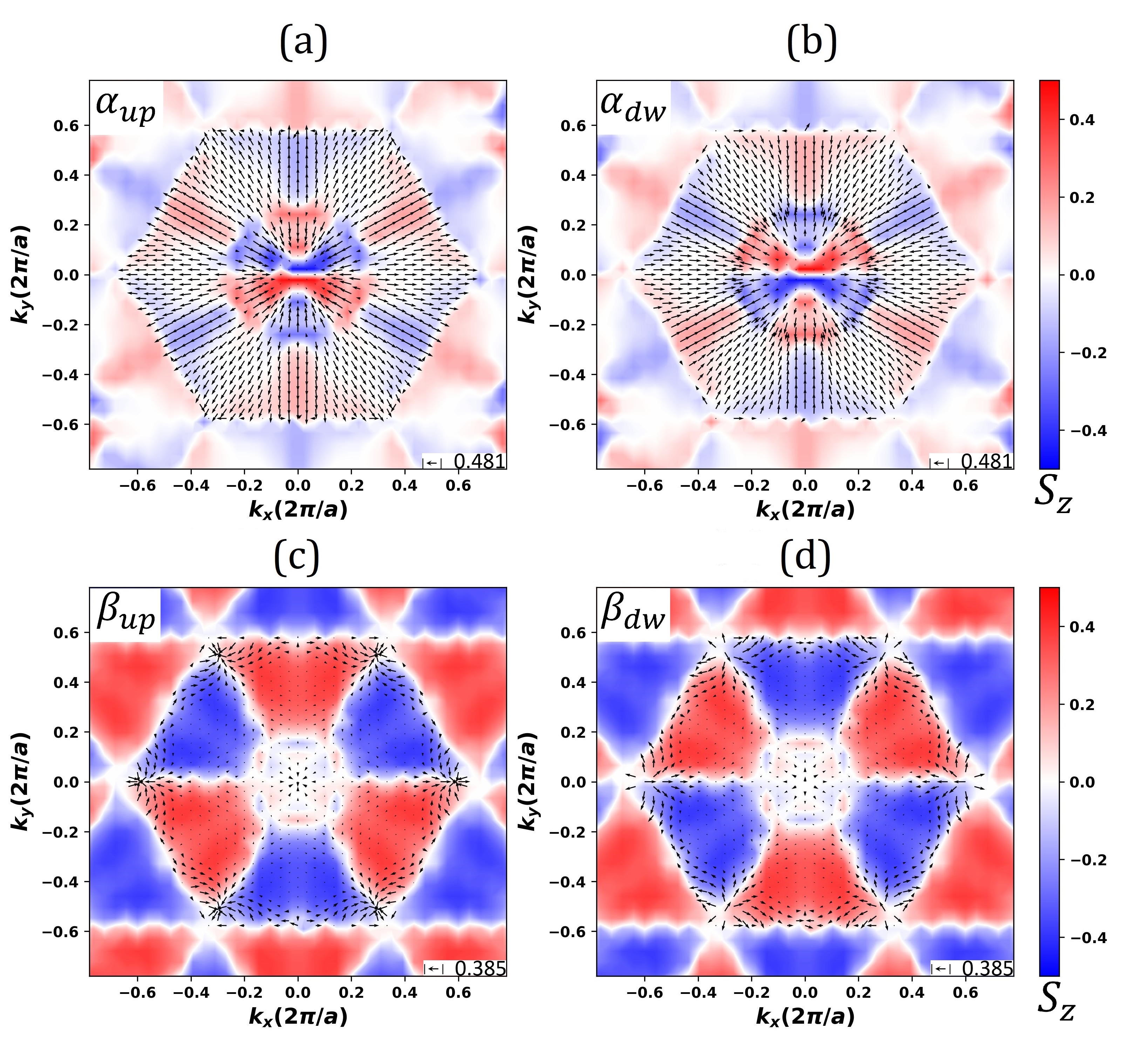}
    \caption{Example of spin textures obtained for the $\alpha$ and $\beta$ valence bands of the $PtTe_2$ monolayer without electric field. The in-plane spin components $(S_{x}, S_{y})$ are shown by black arrows at the $k$-point grids, while the out-of-plane spin texture component is represented by the map of  $S_{z}$ isovalues within the 2D BZ. The length of the arrows represents the magnitude of the in-plane spin components, with the scale provided in the bottom-right corner of the plot. }
    \label{spintexture_without-E}
\end{figure*}

\begin{figure*}
    \centering
    \includegraphics[width=1\linewidth]{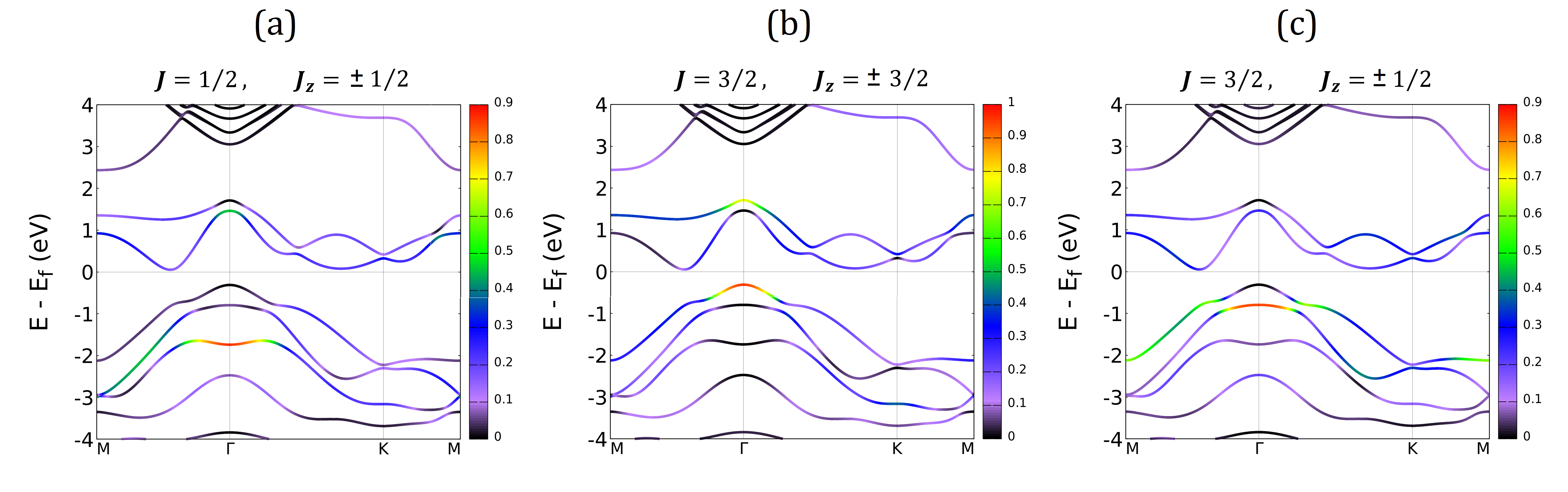}
    \caption{Projected band structure for the $PtTe_2$ monolayer (without electric field). The projections are on the Te-$5p$  atomic states with (a) J = 1/2,  $J_{z}=\pm 1/2$,  (b) J = 3/2,  $J_{z}=\pm 3/2$,  and (c) J = 3/2,  $J_{z}=\pm 1/2$.}
    \label{projected_band}
\end{figure*}

\begin{figure*}[h]
    \centering
    \includegraphics[width=1\linewidth]{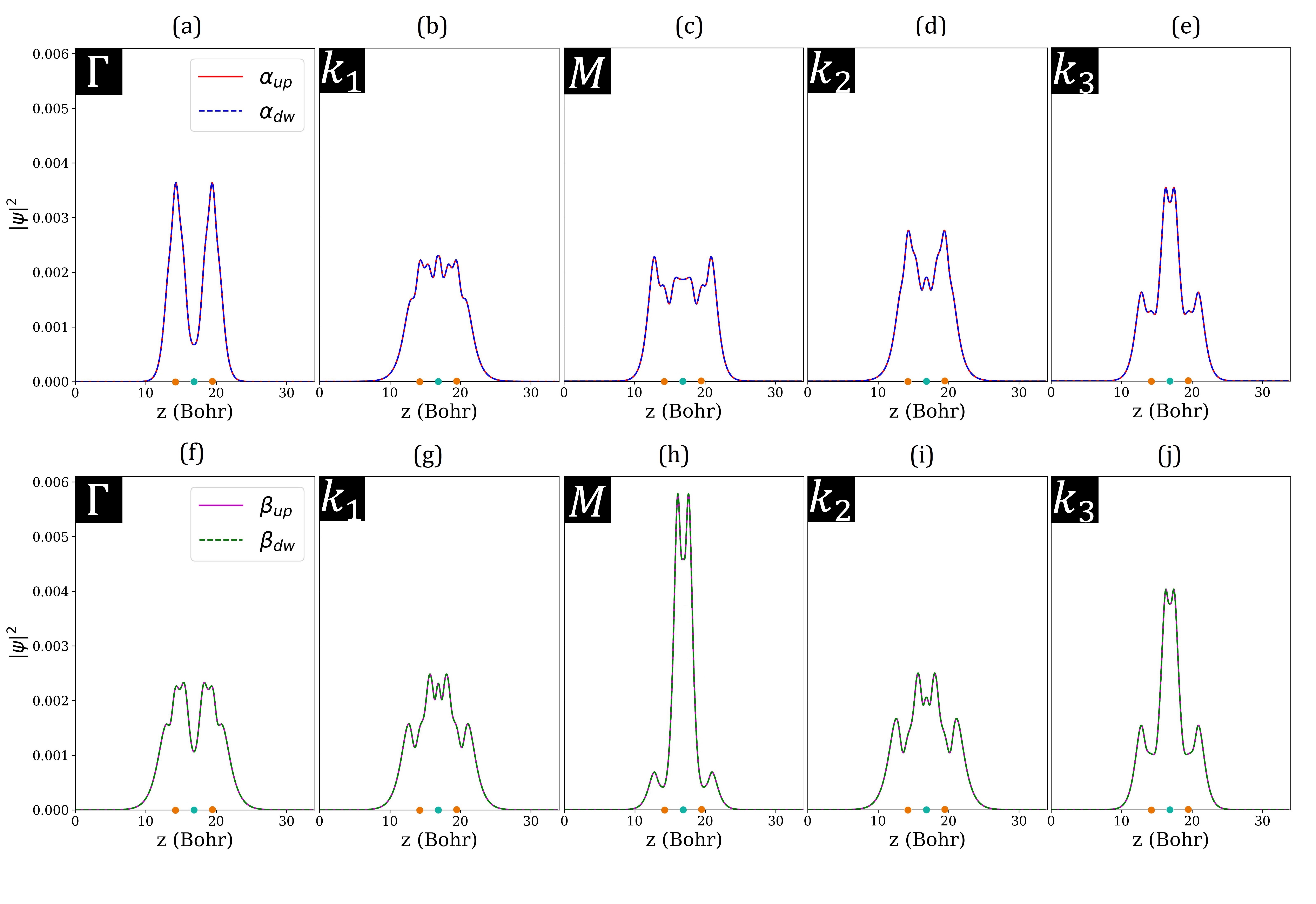}
    \caption{Example of planar-averaged probability distributions (in ${Bohr}^{-3}$) of the $\alpha$- and $\beta$- band eigenstates of the $PtTe_2$ monolayer obtained without electric field at different Brillouin-zone k points. These states correspond to the states used to calculate the spin textures of Fig. \ref{spintexture_without-E} at those {\bf k} points.}
    \label{prob_total_without-E}
\end{figure*}

\begin{figure*}
    \centering
    \includegraphics[width=1\linewidth]{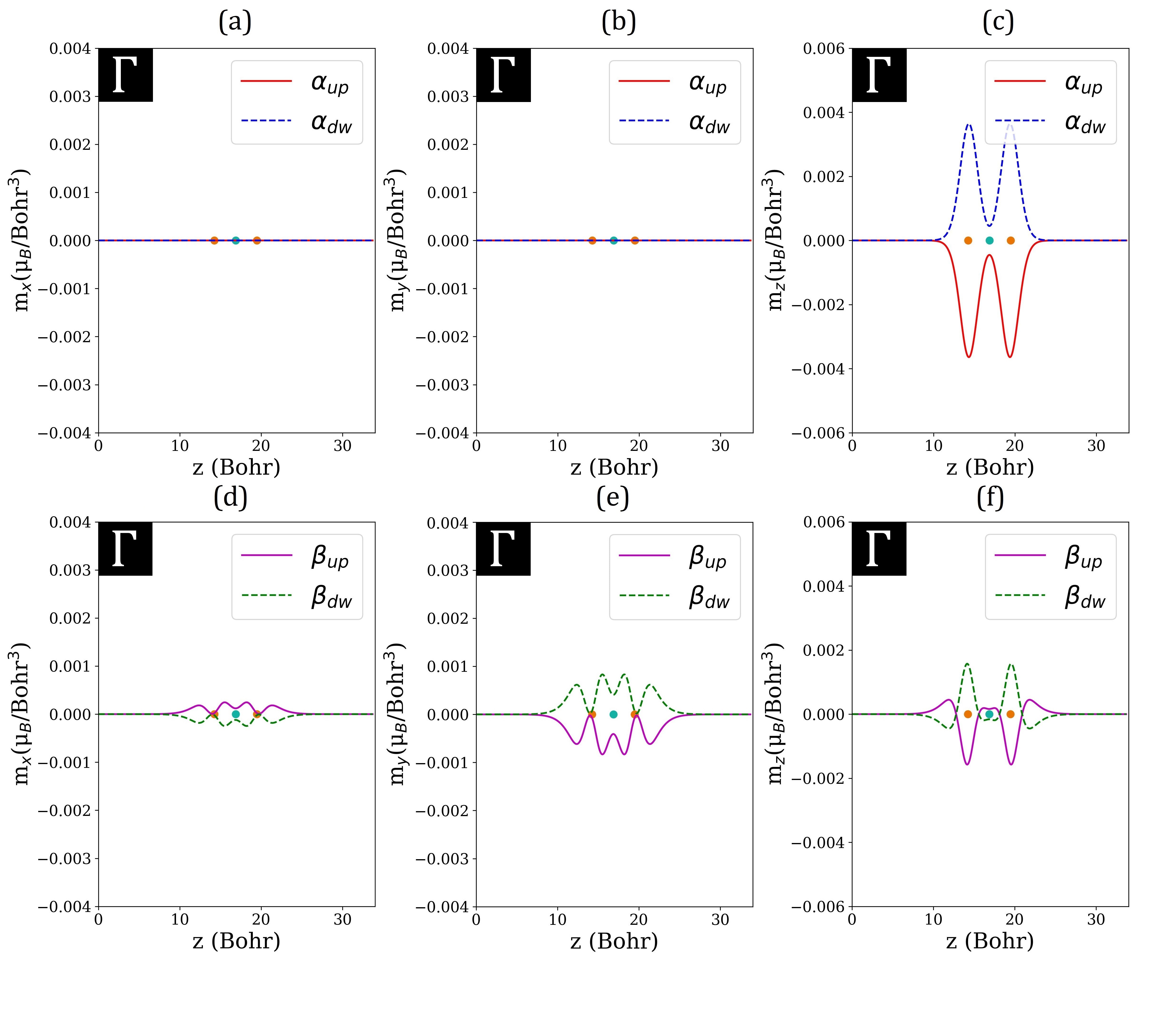}
    \caption{Example of planar-averaged magnetization-density components of the $\alpha$- and $\beta$- band eigenstates of the $PtTe_2$ monolayer obtained without electric field  at the point $\Gamma$. These states have probability densities shown in Figs. \ref{prob_total_without-E}(a) and (f); the spatial integrations of the magnetization-density components of these states produce the spin polarization vectors $(S_z, S_y, S_z)$ at $\Gamma$ of the spin textures shown in Fig. \ref{spintexture_without-E}. }

    \label{mag_G_without-E}
\end{figure*}

\begin{figure*}
    \centering
    \includegraphics[width=1\linewidth]{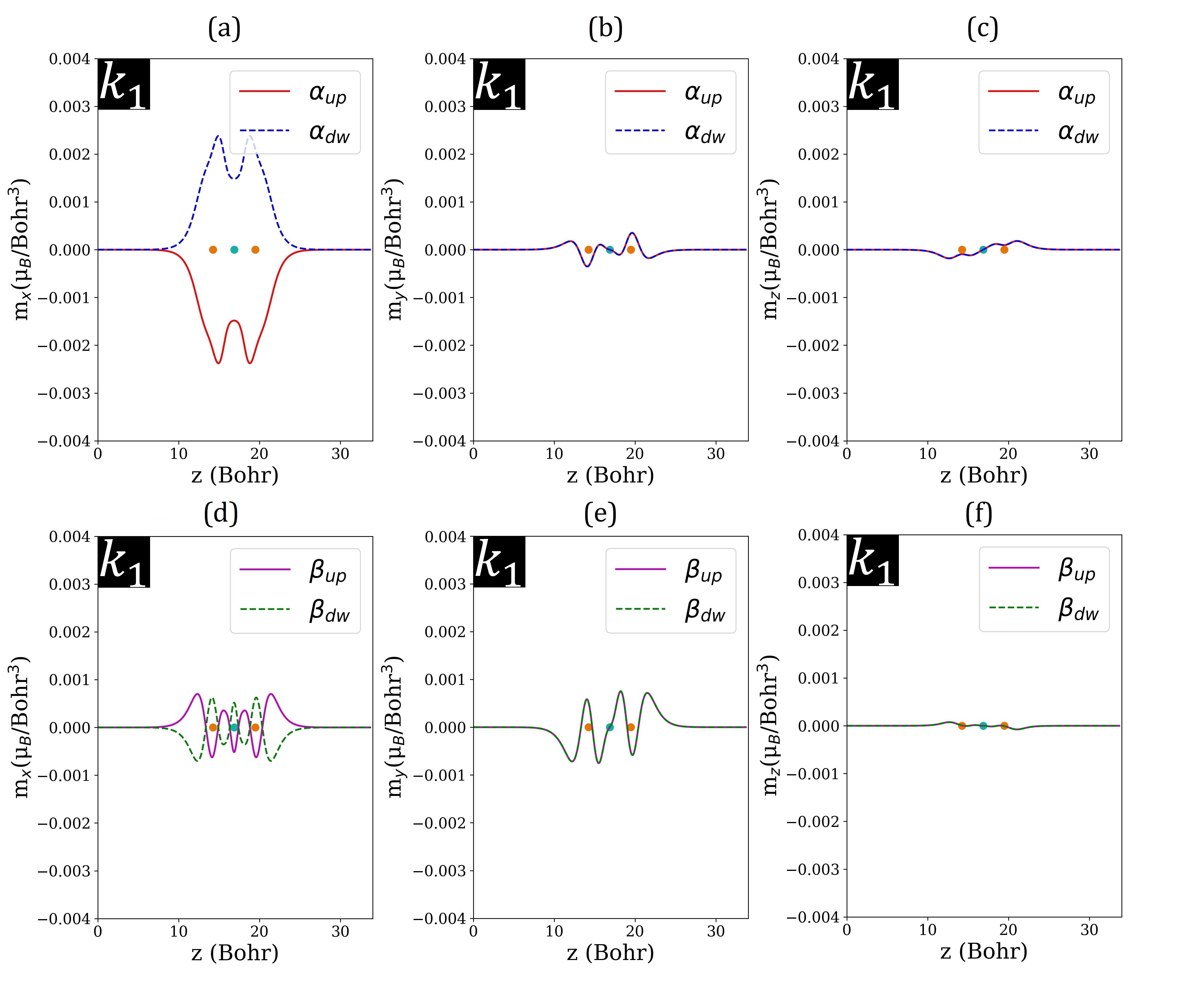}
    \caption{Example of planar-averaged magnetization-density components of $\alpha$- and $\beta$- band eigenstates of the $PtTe_2$ monolayer obtained without electric field  at the point $k_{1}$. These states have probability densities shown in Figs. \ref{prob_total_without-E}(b) and (g); the spatial integrations of the magnetization-density components of these states produce the spin polarization vectors $(S_z, S_y, S_z)$ at  $k_{1}$ of the spin textures shown in Fig. \ref{spintexture_without-E}.  }
    \label{mag_k1_without-E}
\end{figure*}

\begin{figure*}
    \centering
    \includegraphics[width=1\linewidth]{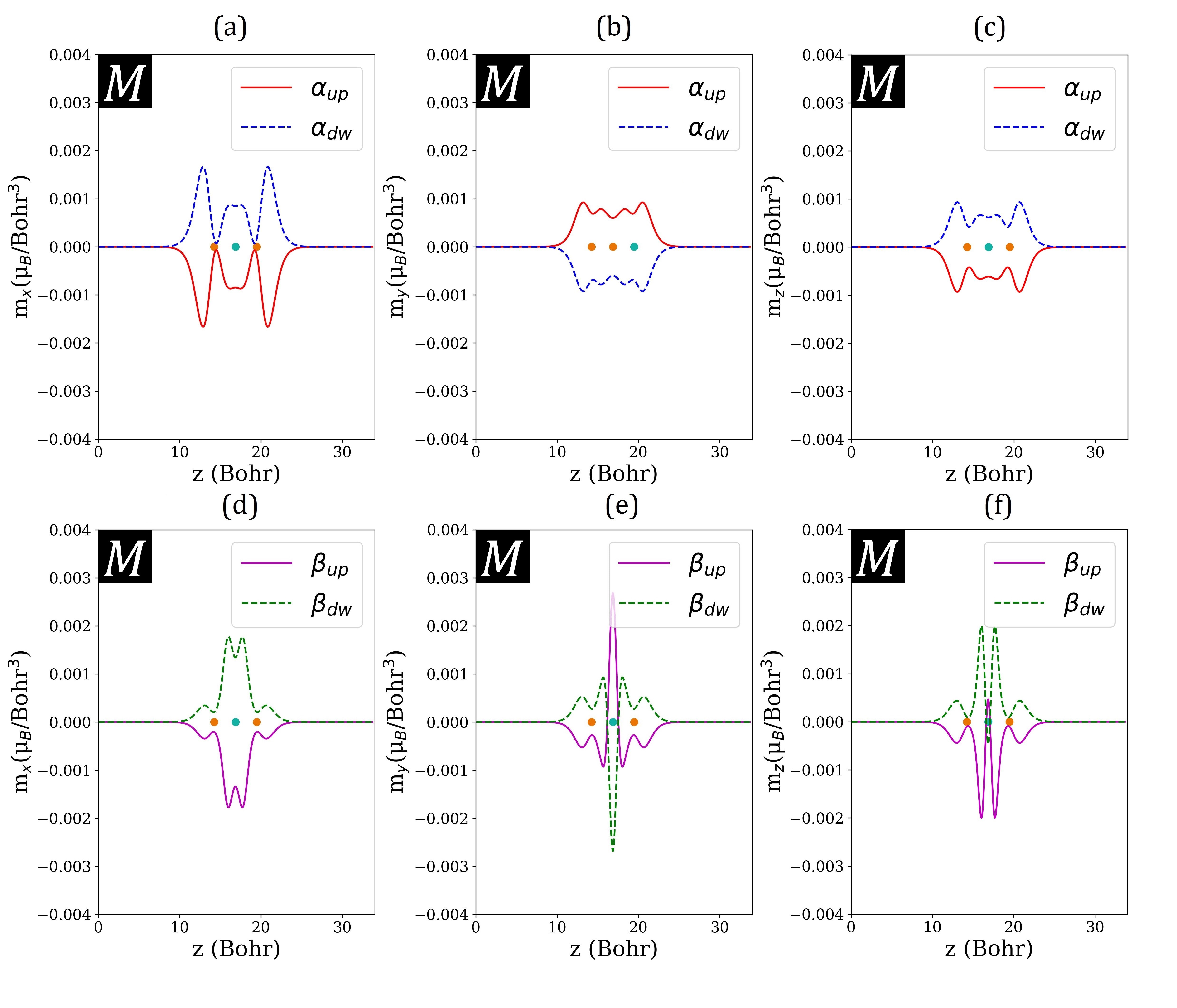}
    \caption{Example of planar-averaged magnetization-density components of $\alpha$- and $\beta$- band eigenstates of the $PtTe_2$ monolayer obtained without electric field  at the point $M$. These states have probability densities shown in Figs. \ref{prob_total_without-E}(c) and (h); the spatial integrations of the magnetization-density components of these states produce the spin polarization vectors $(S_z, S_y, S_z)$ at  $M$ of the spin textures shown in Fig. \ref{spintexture_without-E}.  }
    \label{mag_M_without-E}
\end{figure*}

\begin{figure*}
    \centering
    \includegraphics[width=1\linewidth]{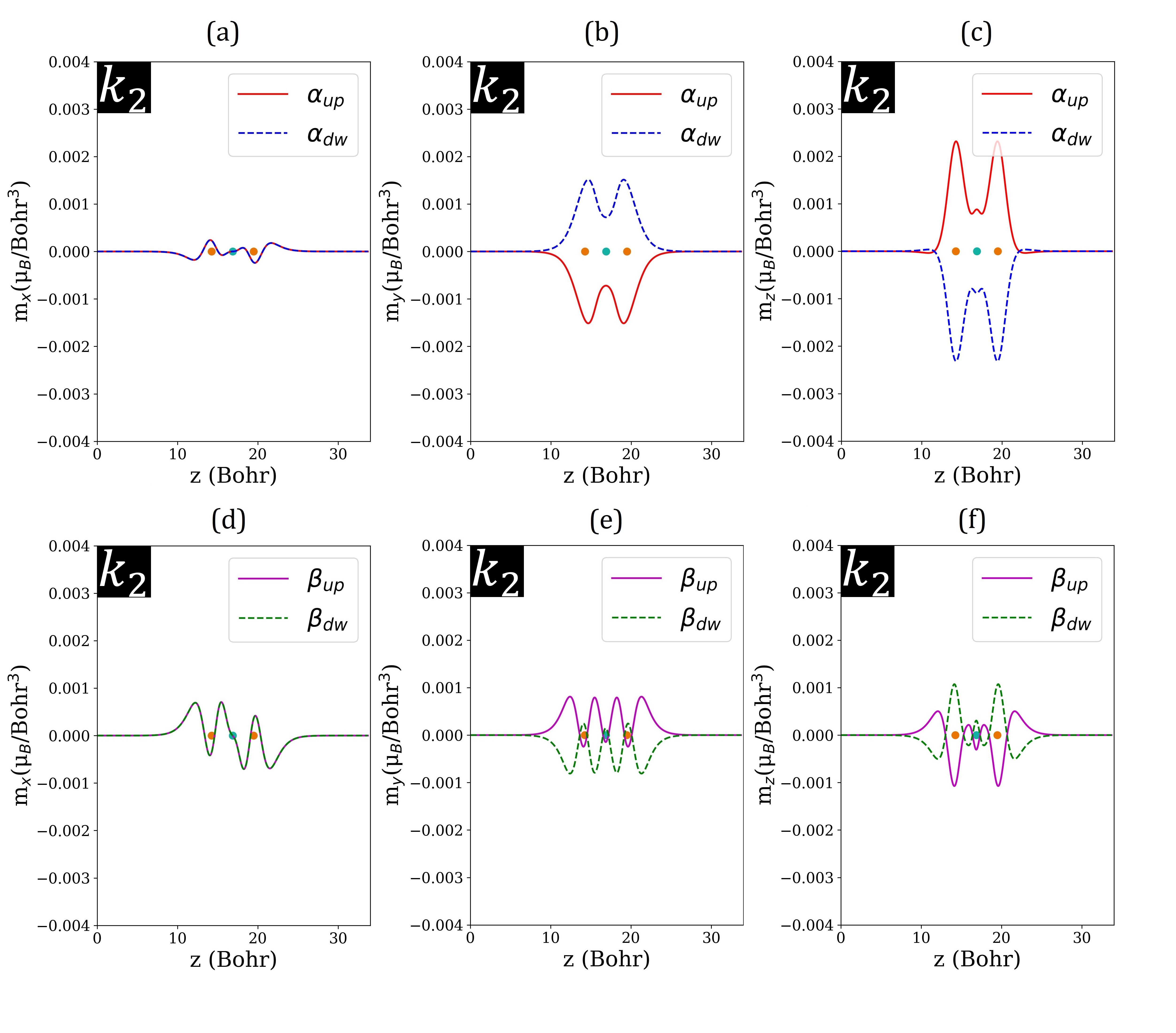}
    \caption{Example of planar-averaged magnetization-density components of $\alpha$- and $\beta$- band eigenstates of the $PtTe_2$ monolayer obtained without electric field at the point $k_{2}$. These states have probability densities shown in Figs. \ref{prob_total_without-E}(d) and (i); the spatial integrations of the magnetization-density components of these states produce the spin polarization vectors $(S_z, S_y, S_z)$ at  $k_{2}$ of the spin textures shown in Fig. \ref{spintexture_without-E}.  }
    \label{mag_k2_without-E}
\end{figure*}

\begin{figure*}
    \centering
    \includegraphics[width=1\linewidth]{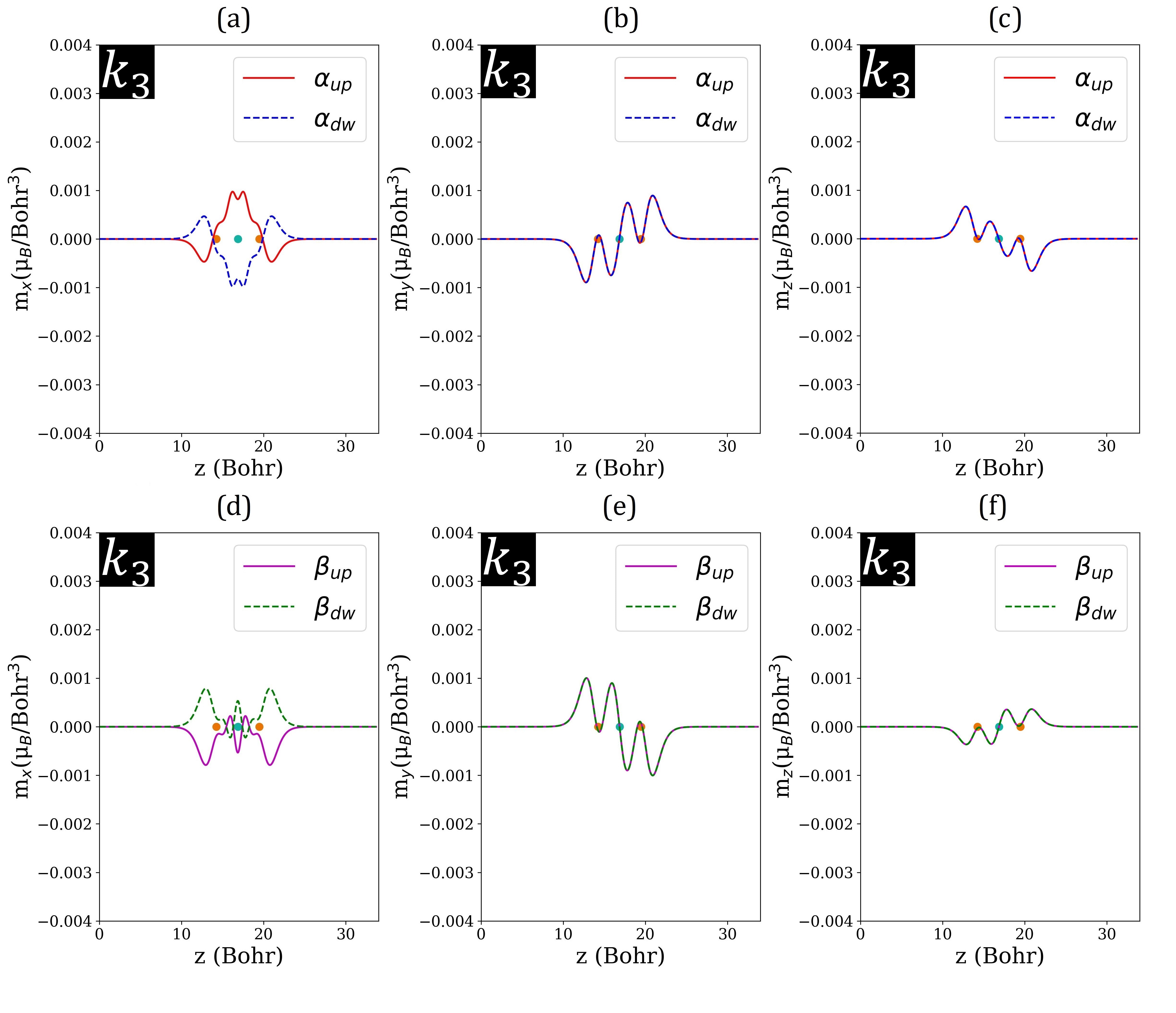}
    \caption{Example of planar-averaged magnetization-density components of $\alpha$- and $\beta$- band eigenstates of the $PtTe_2$ monolayer obtained without electric field at the point $k_{3}$. These states have probability densities shown in Figs. \ref{prob_total_without-E}(e) and (j); the spatial integrations of the magnetization-density components of these states produce the spin polarization vectors $(S_z, S_y, S_z)$ at  $k_{3}$ of the spin textures shown in Fig. \ref{spintexture_without-E}.  }
    \label{mag_k3_without-E}
\end{figure*}

  Fig.~\ref {prob_total_without-E} presents examples of planar-averaged  probability densities of the eigenstates of the doubly-degenerate $\alpha$ and $\beta$ bands of monolayer $PtTe_2$, as obtained with QE in the absence of an external electric field. The probability density distributions are plotted for the five {\bf k} points indicated in Fig.1(c) of the main text. The eigenstates correspond to those used to calculate the spin textures of Fig.~\ref{spintexture_without-E} at those k points.
  As observed in Fig.~\ref{prob_total_without-E}, these eigenstates have inversion-symmetric probability-density distributions, which are identical for the doubly degenerate states. In Figs.~\ref{mag_G_without-E}-\ref{mag_k3_without-E}, we also report
 the planar-averaged magnetization density components of the same eigenstates at those five {\bf k} points.


\section{Influence of the local Rashba effect on the states near $\Gamma$ of the $\alpha$ and $\beta $ bands }

As mentioned in the main text and illustrated in Fig. \ref{projected_band}, the states at and near $\Gamma$ of the upper ($\alpha$) and second-upper ($\beta$) valence band of the PtTe$_2$ monolayer are mainly bonding states made of Te-$5p$ $|J = 3/2, J_{z} = \pm 3/2\rangle$ and  $|J = 3/2, J_{z} = \pm 1/2\rangle $ atomic states, respectively. Near $\Gamma$, such states of the Te layers are subject to the local Rashba effect, caused by the local electric fields $\bm{E}_z^{(l)}$  induced along $z$ on each of the Te layers  $l=l_1,l_2$ by the presence of the other layers, where  $\bm{E}_z^{(l_1)} = - \bm{E}_z^{(l_2)}$.  This effect may be described by the phenomenological Rashba onsite Hamiltonian's $H_{R}^{(l)}= - \lambda ({\bm k}\times \bm{E}_z^{(l)})\cdot \hat{\boldsymbol \sigma} = - \lambda E_z^{(l)} (k_y \hat{\sigma}_x - k_x \hat{\sigma}_y)$, where $\lambda$ is a constant for the Te-$5p$ states. The matrix elements $\langle J, J_{z}|H_{R}^{(l)}|J, J'_z\rangle $ produce a coupling by the Rashba onsite terms between the $|J, J_{z} \rangle$ and  $|J, J'_{z} \rangle $ states and are non zero only for  $\Delta J_z = J_z - J'_z = \pm 1$ ~\cite{Ghosh_FM_bilayer}.

The Te-$5p$ $|J = 3/2, J_{z}\rangle$ atomic states of the ions within a given Te layer have as spin-angular components:  
\begin{equation}
    |J = 3/2, J_{z} = 3/2\rangle = Y_{1}^1 |1/2\rangle,
\end{equation}
\begin{equation}
    |J = 3/2, J_{z} = -3/2\rangle = Y_{-1}^1 |-1/2\rangle, 
\end{equation}
\begin{equation}
    |J = 3/2, J_{z} = 1/2\rangle = \frac{1}{\sqrt3}Y_{1}^1 |-1/2\rangle + \sqrt{\frac{2}{3}}Y_{0}^1 |1/2\rangle,
\end{equation}
\begin{equation}
    |J = 3/2, J_{z} = -1/2\rangle = \sqrt{\frac{2}{3}}Y_{0}^1 |-1/2\rangle + \frac{1}{\sqrt3}Y_{-1}^1 |1/2\rangle,
\end{equation}
where $Y_{m}^l$ are the spherical harmonics,  $Y_{1}^1 = \frac{1}{\sqrt{2}}(p_x+i p_y)$, $Y_{-1}^1=\frac{1}{\sqrt{2}}(p_x-i p_y)$, and 
$Y_{0}^1=p_z$, and $|s_z\rangle$ are the eigenstates of the out-of-plane spin operator  $\frac{1}{2}\hat{\sigma}_z$. The $J_z = \pm 3/2$ states in Eqs. (1-2) are characterized by in-plane $p_{\parallel}$ orbitals, while the $J_z = \pm 1/2$ states of Eqs. (3-4) have both in-plane $p_{\parallel}$ and out-of-plane $p_z$ orbitals. 

The Pauli matrices are: 
\begin{equation} 
\hat{\sigma}_{x} =  
\begin{pmatrix}
0 & 1 \\
1 & 0  
\end{pmatrix},  
\hat{\sigma}_{y} =  
\begin{pmatrix}
0 & -i \\
i & 0  
\end{pmatrix}, 
\hat{\sigma}_{z} =  
\begin{pmatrix}
1 & 0 \\
0 & -1  
\end{pmatrix},    
\end{equation} 
and the $|s_z\rangle$ spin-up and spin-down eigenstates of $\hat{\sigma}_{z}$ in spinor form are:
\begin{equation}
|1/2\rangle = \begin{pmatrix}
1 \\ 0
\end{pmatrix},   
|-1/2\rangle = \begin{pmatrix}
0 \\ 1
\end{pmatrix}.  
\end{equation}
These spin states yield non-vanishing matrix elements $\langle s_z| \hat{\sigma}_{x,y} | s'_z\rangle$ only when $\Delta s_z = s_z - s'_z = \pm 1$, with $\langle 1/2| \hat{\sigma}_x | -1/2 \rangle = \langle -1/2| \hat{\sigma}_x  1/2 \rangle^* = 1 $ and $\langle 1/2| \hat{\sigma}_y | -1/2 \rangle  = \langle -1/2| \hat{\sigma}_y | 1/2 \rangle^* = -i$. 
Consequently, the matrix elements $\langle J=3/2,J_z|\frac{1}{2}\hat{\sigma}_{x,y}|J=3/2, J'_z\rangle$ of the in-plane spin operators $\frac{1}{2}\hat{\sigma}_{x,y}$ are non-zero only for $\Delta J_{z}=J_z-J'_z=\pm 1$; in particular, for $\hat{\sigma}_y$: 
\begin{equation}
 \langle J=3/2, J_z=3/2 | \hat{\sigma}_{y} | J=3/2, J'_z= 1/2\rangle =  - \frac{i}{\sqrt{3}},
 \end{equation}
\begin{equation}
 \langle J=3/2, J_z=-3/2 | \hat{\sigma}_{y} | J=3/2, J'_z= -1/2\rangle = \frac{i}{\sqrt{3}},
 \end{equation}
\begin{equation}
\langle J=3/2, J_z=1/2|\hat{\sigma}_{y}|J=3/2, J'_z=-1/2\rangle = -\frac{2}{3}i,
\end{equation}
and the corresponding $\langle J,J'_z|\hat{\sigma}_y|J, J_z\rangle$ matrix elements are the complex conjugates. 

In the limit of no Rashba term, the states of the $\alpha$ and $\beta$ bands at and near $\Gamma$ correspond essentially to the bonding states formed using the same component $|J, J_{z} \rangle (l_1)$ and $|J, J_{z} \rangle (l_2)$ of Eqs. (1),(2),(3), and (4) on the Te layer $l_1$ and $l_2$, where $J_z = +3/2, -3/2 $ for the doubly-degenerate states of the $\alpha$ band and $J_z = +1/2, -1/2 $ for  the doubly-degenerate states of the $\beta$ band. These bonding states exhibit 
spin polarization densities $\psi_{J,J_z}^{+}(\textbf{r})\hat{\sigma}_{\alpha}\psi_{J,J_z}(\textbf{r}) $ that are globally inversion symmetric between the two layers and that cancel (are opposite) at any {\bf r} point for the $J_z = 3/2$ and $-3/2$ states of the $\alpha$ band,  and for the $J_z = 1/2$ and $-1/2$ states of the $\beta$ band --- consistent with the observations at $\Gamma$ in our study. 

In the other limit of no interlayer bonding potential, the states of the lowest-energy bands near $\Gamma$ with dominant spin-orbital character given by Eq.(1), (2), (3) or, (4) can be expected to include some additional component due to the Rashba coupling.  For any small ${\bm k}$ around $\Gamma$ in the 2D BZ, if one  chooses the reference $k_x$ axis along that ${\bm k}$, the phenomenological Rashba onsite terms at that ${\bm k}$ may be written as $H_R^{(l)}= \lambda k_{x} E_z^{(l)} \hat{\sigma}_y$. The corresponding relevant (non-vanishing) Rashba onsite matrix elements are:  
\begin{equation} 
\langle J= 3/2, J_{z} = 3/2 |\lambda k_{x} E_z \hat{\sigma}_y | J= 3/2, J'_{z} =1/2 \rangle =  - \frac{i}{\sqrt{3}} \lambda k_{x} E_z, 
\end{equation} 
\begin{equation}
\langle J= 3/2, J_{z} = - 3/2 |\lambda k_{x} E_z \hat{\sigma}_y | J= 3/2, J'_{z} = -1/2 \rangle =  \frac{i}{\sqrt{3}} \lambda k_{x} E_z,
\end{equation} 
\begin{equation}
\langle J= 3/2, J_{z} = 1/2 |\lambda k_{x} E_z \hat{\sigma}_y | J= 3/2, J'_{z} = -1/2 \rangle = - \frac{2}{3} i \lambda k_{x} E_z,
\end{equation} 
with the $\langle J,J'_z|\hat{\sigma}_y|J, J_z\rangle$ matrix elements being their complex conjugates. 

For the $J_z= \pm 3/2$ band states on layer $l_1$, the presence of the Rashba terms at that ${\bm k}$ leads to a second-order energy correction with associated perturbed states \cite{Ghosh_FM_bilayer}: 
\begin{equation}
| J= 3/2, J_{z} = \pm 3/2 \rangle'(l_1) = | J= 3/2, J_{z} = \pm 3/2 \rangle(l_1) \mp \frac{i}{\sqrt{3}} \frac{\lambda k_{x} E_z^{(l_1)}}{\Delta E_{\alpha \beta}} | J= 3/2, J_{z} = \pm 1/2 \rangle(l_1) 
\end{equation} 
and analogously, for the $J_z= \pm 3/2$ band states on layer $l_2$, the perturbed states read: 
\begin{equation}
| J= 3/2, J_{z} = \pm 3/2 \rangle'(l_2) = | J= 3/2, J_{z} = \pm 3/2 \rangle(l_2) \pm   \frac{i}{\sqrt{3}} \frac{\lambda k_{x} E_z^{(l_1)}}{\Delta E_{\alpha \beta}} | J= 3/2, J_{z} = \pm 1/2 \rangle(l_2),  
\end{equation} 
where $\Delta E_{\alpha \beta}$ represents the energy difference between the  $| J= 3/2, J_{z} = \pm 3/2 \rangle $ and  $| J= 3/2, J_{z} = \pm 1/2 \rangle $ states of each layer. The Rashba interaction can be expected thus to induce in the $J_z = \pm 3/2$ states of the $\alpha$ band a small $p_z$ component (as $k_x$ is small) through the addition of a small $| J= 3/2, J_{z} = \pm 1/2 \rangle $ component.  It is important to note that the corrections for the states  $J_z=3/2$ $(-3/2)$ on layer $l_1$ and layer $l_2$ have opposite signs in Eqs. (13) and (14), because of the opposite local electric fields on the two layers. Considering then the interlayer potential in the limit of a very small $k_x$ and forming the bonding combination (for the $\alpha$ band) of these perturbed states $J_z=3/2$ $(-3/2)$ on the two layers results in a perturbed bonding state $J_z=3/2$ $(-3/2)$ with an induced $m_y(\textbf{r})$ magnetization-density component that is globally inversion antisymmetric (opposite) between the two layers (because of the opposite locale electric field on the two layers) and that is the same for the perturbed $J_z=3/2$ and $J_z= -3/2$ bonding states. Summing over the two states yields a non-vanishing total magnetization density component along the $y$ direction for the $\alpha$ band that is of opposite sign between the two layers, consistent with the observation in our study. We note that the total magnetization density is an invariant, i.e., independent of the choice of the two orthogonal states in the degenerate subspace. 

For the  $J_z= \pm 1/2$ band states on layer $l_1$, in the limit of no interlayer potential, the presence of the Rashba onsite terms at that ${\bm k}$ can be expected to generate a first-order energy correction through the induced Rashba coupling between the degenerate $J_z= 1/2$ and  $J_z= -1/2$ states on that layer. Diagonalizing the corresponding 2x2 matrix of $H_{R}^{(l_1)}$ in that subspace, yields as eigenstate with lowest energy on layer $l_1$: 

\begin{equation}
\psi_{l_1}  =  \frac{1}{\sqrt{2}} | J= 3/2, J_{z} = 1/2 \rangle(l_1) - \frac{i}{\sqrt{2}} \frac{E_z^{(l_1)}}{|E_z^{(l_1)}|}| J= 3/2, J_{z} = - 1/2 \rangle(l_1);   
\end{equation} 
analogously, the Rashba coupling $H_{R}^{(l_2)}$ yields as eigenstate with the same lowest energy on layer $l_2$: 

\begin{equation}
\psi_{l_2}  =  \frac{1}{\sqrt{2}} | J= 3/2, J_{z} = 1/2 \rangle(l_2) + \frac{i}{\sqrt{2}} \frac{E_z^{(l_1)}}{|E_z^{(l_1)}|}| J= 3/2, J_{z} = - 1/2 \rangle(l_2).   
\end{equation} 
These states of the $\beta$ band located on layer $l_1$ and $l_2$, respectively, include $J_z=-1/2$ components of opposite sign in Eqs. (15) and (16), due to the opposite local electric fields on the two layers. This gives rise to non-vanishing spin polarization vectors along the $y$ axis, which are oriented in opposite directions on layer $l_1$ and $l_2$. The corresponding magnetization density components along the $y$ axis for these states have planar averages which are opposite between the two layers and read: $\bar{m}_y(l_1) = \frac{E_z^{(l_1)}}{|E_z^{(l_1)}|} [\frac{1}{6} [\bar{|p_x|}^2(l_1) - \bar{|p_y|}^2(l_1)] - \frac{2}{3} \bar{|p_z|}^2(l_1)]$ and   $\bar{m}_y(l_2) = \frac{E_z^{(l_1)}}{|E_z^{(l_1)}|} [- \frac{1}{6} [\bar{|p_x|}^2(l_2) - \bar{|p_y|}^2(l_2)] + \frac{2}{3} \bar{|p_z|}^2(l_2)]$. 
Summing over the two states yields for the $\beta$ band a non-vanishing total magnetization density component along the $y$ direction $\bar{m}_y(l_1) + \bar{m}_y(l_2)$ that has opposite signs on the two layers, consistent with the general trend seen around $\Gamma$ in our study. Close to $\Gamma$, when the interlayer potential becomes significant compared to the Rashba term, the situation can be expected to be intermediate between that of the fully segregated states with opposite magnetization densities  $\bar{m}_y(l_1)$ and  $\bar{m}_y(l_2)$ on the two layers and that of the bonding states $ | J= 3/2, J_{z} = + 1/2 \rangle$ and $ | J= 3/2, J_{z} = - 1/2 \rangle$ exhibiting  globally inversion symmetric magnetization densities that are opposite (cancel) at any {\bf r}. In such intermediate situation, one may expect thus intermediate magnetization densities for the segregated states with non-vanishing components along $y$ verifying: $\bar{m}_y^{up}(-z) = - \bar{m}_y^{dw}(z)$ (because of the $T \cdot I$ symmetry at any {\bf k} point), but which are more globally symmetric than the $\bar{m}_y(l_1)$ and $\bar{m}_y(l_2)$ of the fully segregated states. Furthermore, in the intermediate situation, one may expect the total magnetization to be also intermediate, and hence to have a profile similar to $\bar{m}_y(l_1) + \bar{m}_y(l_2)$ with an amplitude that decreases with decreasing $k_x$. These various features are generally consistent with the observations in our study.

\bibliography{supplementary}